# Thermoelectric Properties of Benzothieno-Benzothiophene Self-Assembled Monolayers in Molecular Junctions.

Sergio Gonzalez-Casal,[1] Rémy Jouclas,[2] Imane Arbouch,[3] Yves Geerts,[2,4] Colin van Dyck,[5] Jérôme Cornil,[3] and Dominique Vuillaume.[1*]

1) *Institute for Electronics Microelectronics and Nanotechnology (IEMN), CNRS, Av. Poincaré, Villeneuve d'Ascq, France.*
2) *Laboratory of Polymer Chemistry, Université Libre de Bruxelles, Bd. du triomphe, Bruxelles, Belgium*
3) *Laboratory for Chemistry of Novel Materials, University of Mons, Belgium.*
4) *International Solvay Institutes of Physics and Chemistry, Université Libre de Bruxelles, Bd. du Triomphe, Bruxelles, Belgium*
5) *Theoretical Chemical Physics group, University of Mons, Mons, Belgium.*

** Corresponding author:* dominique.vuillaume@iemn.fr

**Abstract.** We report a combined experimental (C-AFM and SThM) and theoretical (DFT) study of the thermoelectric properties of molecular junctions made of self-assembled monolayers on Au of thiolated benzothieno-benzothiophene (BTBT) and alkylated BTBT derivatives (C8-BTBT-C8). We measure the thermal conductance per molecule at 15 pW/K and 8.8 pW/K, respectively, among the lowest values for molecular junctions so far reported (10-50 pW/K). The lower thermal conductance for $C_8$-BTBT-$C_8$ is consistent with two interfacial thermal resistances introduced by the alkyl chains, which reduce the phononic thermal transport in the molecular junction. The Seebeck coefficients are 36 µV/K and 245 µV/K, respectively, the latter due to the weak coupling of the core BTBT with the electrodes. We deduce a thermoelectric figure of merit *ZT* up to ≈ $10^{-4}$ for the BTBT molecular junctions at 300K, on a par with the values reported for archetype molecular junctions (olygo(phenylene ethynylene) derivatives).

Molecular junctions (MJs) have been suggested as efficient thermoelectric devices at the nanoscale.[1] Due to their nanoscopic scale and quantum behavior, the classical Fourier's law is broken down[2] and among other deviations, the thermal conductivity of MJs is size-dependent or equivalently the thermal conductance no longer scales as 1/L (with L the molecule length) but is almost constant or varies as $1/L^{0.64}$.[3-9] More research is definitely required to explore molecular thermoelectricity and to develop efficient thermal management strategies for molecular nano-devices.[10] The complete study of the thermoelectric properties of MJs requires measuring the electrical conductance $G_{el}$, the Seebeck coefficient $S$ (thermopower) and the thermal conductance $G_{th}$ (equivalently, the conductances translate to the electrical conductivity $\sigma$ and to the thermal conductivity $\kappa$ considering the *ad-hoc* geometrical factors of the measured devices).[11] The experimental determination of the thermal conductance of molecular junctions is relatively scarce compared to the first two factors (see reviews in Refs. 9, 12, 13), especially considering studies at the nanoscale in devices featuring self-assembled monolayers (SAMs) or single molecules. Moreover, the results reported so far concern only MJs based on alkanethiols and oligo(phenylene ethynylene) (OPE). Meier et al.[7] studied the thermal conductance of SAM-based MJs (Au-alkanethiol-Au) using scanning thermal microscopy (SThM) and reported a thermal conductance (per molecule) in the range of 10-30 pW/K decreasing as $L^{0.64}$ for alkyl chain length between 2 and 18 carbon atoms. More recently, SThM-based experiments on single molecule MJs measured a thermal conductance in the range of 15-30 pW/K for Au-alkanedithiol-Au (almost constant for 2 to 10 carbon atoms),[8] ≈37 pW/K (for 8 carbon atoms)[14] and ≈23-24 pW/K for OPE-based MJs.[11, 14]

Here, we report the thermoelectric properties of SAM-based MJs made of benzothieno-benzothiophene (BTBT) thiolated derivatives. BTBT derivatives are interesting molecules for several applications in electronics such as transistors owing to their high charge mobility.[15, 16] We previously compared the thermal conductivity (measured by SThM) of polycrystalline thin films (40-400 nm) of BTBT and alkylated BTBT derivatives (octyl chains at the α and ω positions of the BTBT core, $C_8$-BTBT-$C_8$).[17] From a combined SThM and computational study, we unveiled that the thermal conductivity of the BTBT films is larger than that of the alkylated BTBT (0.63 W $m^{-1}$ $K^{-1}$ vs. 0.25 W $m^{-1}$ $K^{-1}$) because the alkyl chains strongly localize the phonon modes in the BTBT layers. In the present work, we synthesized the thiol functionalized derivatives of the same molecules to form SAMs on Au electrodes and characterized the thermal and electrical conductances of these SAM-based MJs by SThM and conductive-AFM (C-AFM). The thermal conductivity of the BTBT SAM is measured at $\kappa_{SAM(BTBT)} = 0.46 \pm 0.27$ W $m^{-1}$ $K^{-1}$ and $\kappa_{SAM(C8\text{-}BTBT\text{-}C8)} = 0.27 \pm 0.16$ W $m^{-1}$ $K^{-1}$ (*i.e.*

a thermal conductance of the SAMs $G_{th,SAM(BTBT)}$ = 37±21 nW/K and $G_{th,SAM(C8\text{-}BTBT\text{-}C8)}$ = 22±13 nW/K. From these values, we estimated the thermal conductance per molecule at 15 pW/K and 8.8 pW/K, respectively, which is slightly lower than that recently measured for alkanethiol and OPE MJs (*vide supra*), and makes BTBT derivatives attractive candidates to optimize the thermoelectric figure of merit. The mean electrical conductance (per molecule) is estimated at $2.7 \times 10^{-10}$ S ($3.5 \times 10^{-6} G_0$, with $G_0$ the quantum of conductance $G_0 = 2e^2/h$ = 77.5 µS) for the BTBT molecules and is not measurable for the $C_8$-BTBT-$C_8$ molecules. Electron transport (ET) properties through the BTBT SAMs reveal a broad distribution of the conductance and of the energy of the molecular orbital involved in the ET. These features are explained by DFT calculations considering several configurations of the molecules in the SAM (tilt and twist angles, packing density). We also calculated at the DFT level the Seebeck coefficients to be $S_{BTBT} \approx 36$ µV/K and $S_{C8\text{-}BTBT\text{-}C8} \approx 245$ µV/K.

Two molecules were specifically synthesized with a very different alkyl chain length (see details of the synthesis routes and characterization in the Supporting Information): HS-C-BTBT and HS-$C_8$-BTBT-C8, where BTBT stands for [1]benzothieno[3,2-b][1]benzothiophene (Fig. 1). The synthesis of the BTBT analog bearing a thiol anchor for SAM experiments was first considered by introducing the thiol group directly on the BTBT core. However, this method would probably have led to an unstable material sensitive to oxidation into sulfoxide or sulfone, as it has been previously described by some of us for 2,7-dithio-BTBT.[18] We thus considered synthesizing HS-C-BTBT in 3 steps (Fig. 1) starting from the regioselective formylation of BTBT **1** at the 2$^{nd}$ position, as previously described by Košata et. al.[19] using modified conditions inspired by a patent from Etori et. al.[20] These conditions, although enabling the increase of the regioselectivity of the formylation on the 2$^{nd}$ position over the 4$^{th}$ from 2:1 to 7:1 with respect to Košata's procedure, and reaching a conversion ratio of BTBT of about 89%, gave a yield of only 24% due to the difficulty to isolate all the products by column chromatography. The reduction of aldehyde **2** with sodium tetraborohydride was performed according to a patent of Kawakami and Yamaguchi[21] in good yields. Finally, substitution of the hydroxy group by thiourea followed by its hydrolysis to thiol were performed following the procedure of Cho et. al.,[22] affording HS-C-BTBT with a moderate through honorable overall yield of 12% considering the propensity of the compound to dimerize.

The synthesis of HS-$C_8$-BTBT-$C_8$ was performed by introducing end-brominated octyl chain on monooctyl-BTBT **5** in view of introducing the thiol group at the final steps. Synthesis of C8-BTBT **5** was carried out using the conditions previously described,[23] and the introduction of the η-bromooctyl chain was performed by Friedel-Crafts reaction using the corresponding carboxylic acids in good yields. Reduction of the corresponding ketones **6** keeping safe the bromine atom was performed using the conditions previously published[24] involving the use of $AlCl_3$ and $LiAlH_4$ 1M in diethyl ether to afford compound **7** in good yields. Finally, the same substitution-hydrolysis sequence as used for the synthesis of HS-C-BTBT adapted from the work of Cho et. al.[22] allowed us to obtain HS-$C_8$-BTBT-$C_8$ in a moderate overall yield of 26%.

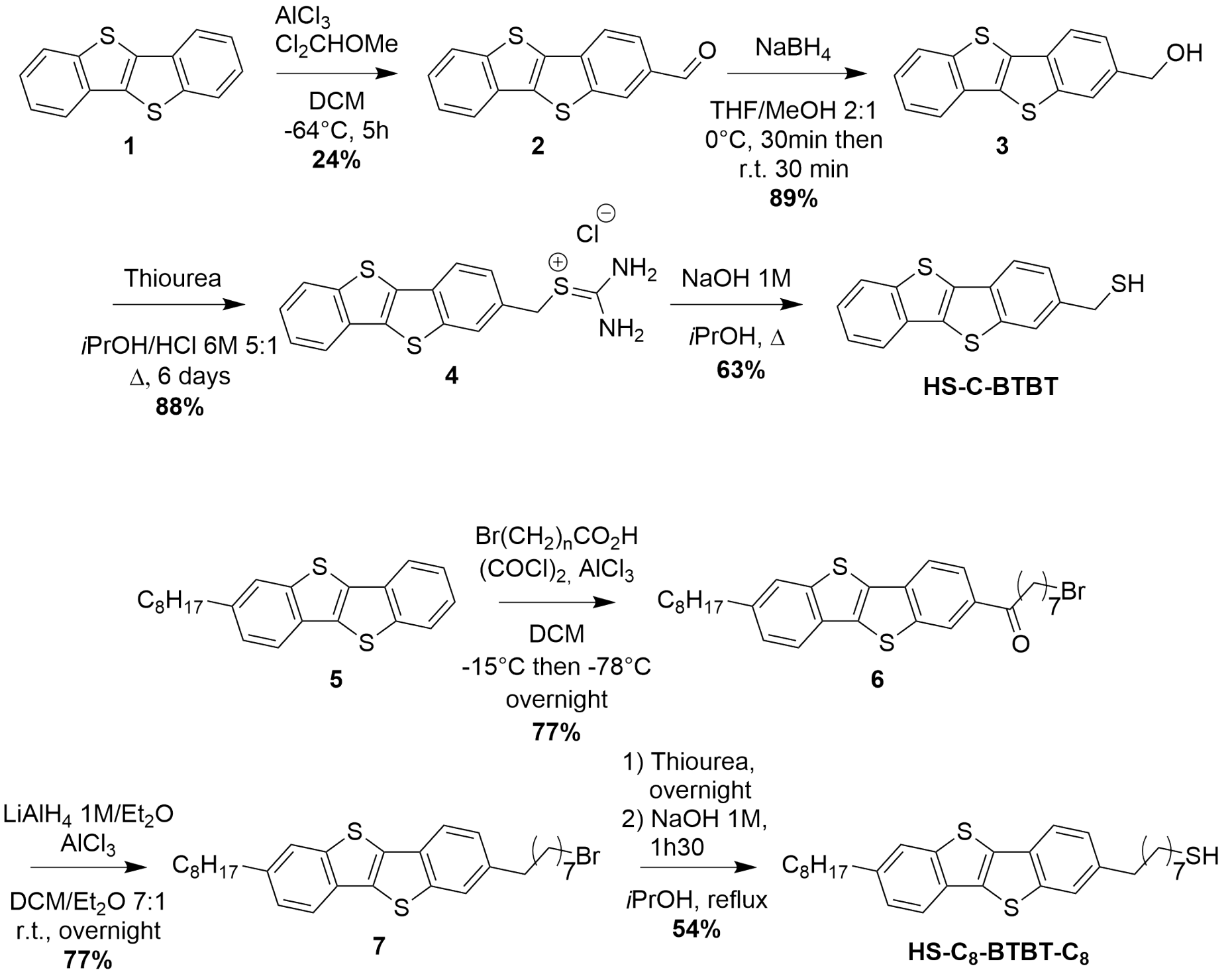

***Figure 1**. Schemes of the synthesis of HS-BTBT and HS-$C_8$-BTBT-$C_8$.*

The SAMs were formed in solution on ultra-flat template stripped Au surface ($^{TS}$Au) and their thicknesses were measured by ellipsometry at 0.9±0.2 nm and 2.6±0.2 nm, in good agreement with the values deduced from geometry optimization of the molecule/$^{TS}$Au interfaces (see details in the Supporting Information). The topographic AFM images (Fig. S1 in the Supporting Information) show that the SAMs are free of gross defects with a rms roughness of 0.6-0.7 nm close to our naked $^{TS}$Au substrate (*ca.* 0.4 nm).[25]

The thermal conductivity of the SAMs was measured by the null-point scanning thermal microscope (NP-SThM) method.[26] In brief (see details in the Supporting Information), this is a differential method measuring the tip temperature jump ($T_{NC}$-$T_C$) when the tip enters in contact with the surface sample, $T_{NC}$ and $T_C$ being the tip temperature just before and after the contact.

Starting from a remote position, the heated tip is approached to the surface and the measured tip temperature starts decreasing slowly because the heat transfer through the air gap is increased. At contact, the sudden jump from $T_{NC}$ to $T_C$ is due to the additional heat flux passing through the SAM/tip contact. This approach allows removing the parasitic contributions (*e.g.* air thermal conduction, radiation). Figure 2 shows typical tip temperature *versus* tip vertical displacement (25 $_{ttip}$-*z* traces per sample) measured on the $^{TS}$Au-S-C-BTBT and $^{TS}$Au-S-$C_8$-BTBT-$C_8$ SAMs. These curves were measured for several heat fluxes passing through the $^{TS}$Au-SAM/tip junctions (*i.e.* by increasing the heating of the tip by increasing the voltage, $V_{WB}$, applied to the Wheatstone bridge in which the SThM tip is inserted, the substrate being at ambient temperature, see the Supporting Information).

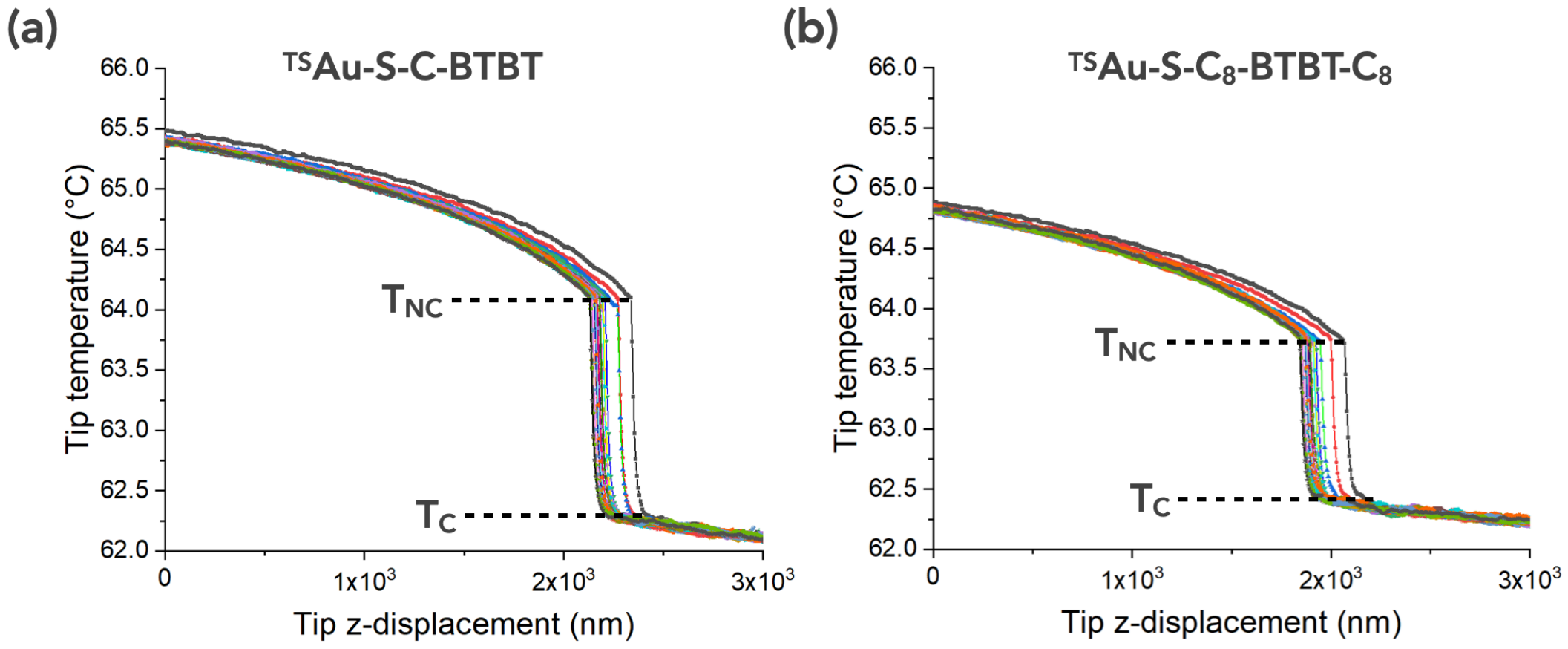

***Figure 2***. *Typical tip temperature vs. tip-surface distance measured at $V_{WB}$=1.1 V during the approach z-scan (0 corresponds to the tip retracted), 25 traces acquired sequentially at the same location on the SAM:* ***(a)*** *on the $^{TS}$Au-S-C-BTBT and* ***(b)*** *on the $^{TS}$Au-S-$C_8$-BTBT-$C_8$ SAMs.*

Figure 3 shows the $T_C$ *vs.* ($T_{NC}$-$T_C$) curves, at several heat fluxes, for the two samples. The thermal conductivity of the SAM/Au samples, $\kappa_{SAM/Au}$, is calculated from the slope of these curves, according to the relationship:[26]

$$T_C - T_{amb} = \left(\alpha \frac{1}{\kappa_{SAM/Au}} + \beta\right)(T_{NC} - T_C) \quad (1)$$

where $\alpha$ and $\beta$ are calibration parameters dependent on the tip and equipment. They were systematically measured for all the data shown in Fig. 3 (see the Supporting Information) and $T_{amb}$ is the room temperature (22.5°C in our air-conditioned laboratory). Due to the very weak thickness of the SAMs, the high thermal conductivity of the Au electrode (318 W m$^{-1}$ K$^{-1}$) contributes to these measured values. The thermal conductivity of the SAM, $\kappa_{SAM}$, is obtained by correcting this measured value from the Au electrode contribution using the Dryden model,[27] following the same approach as in our previous work on the thermal conductivity of very thin organic films (<10 nm) of PEDOT:OTf deposited on Au electrode[28] (see also details in the Supporting Information, Eq. S1). The values of $\kappa_{SAM}$ are shown on Fig. 3-c for measurements done on three distinct zones (randomly chosen) for each sample (also summarized in Table 1 with the $\kappa_{SAM/Au}$ values). On average, we got $\kappa_{SAM(BTBT)}$ = 0.46 ± 0.27 W m$^{-1}$ K$^{-1}$ and $\kappa_{SAM(C8\text{-}BTBT\text{-}C8)}$ = 0.27 ± 0.16 W m$^{-1}$ K$^{-1}$. Equivalently, the thermal conductance of the MJs, $G_{th,SAM}$, is given by $G_{th,SAM}=4r_{th}\kappa_{SAM}$ at the thermal constriction between tip and SAM[29] ($r_{th}$ is the thermal contact radius ≈ 20 nm, see the Supporting Information, Eq. S2). The mean values are $G_{th,SAM(BTBT)}$ = 37±21 nW/K and $G_{th,SAM(C8\text{-}BTBT\text{-}C8)}$ = 22±13 nW/K.

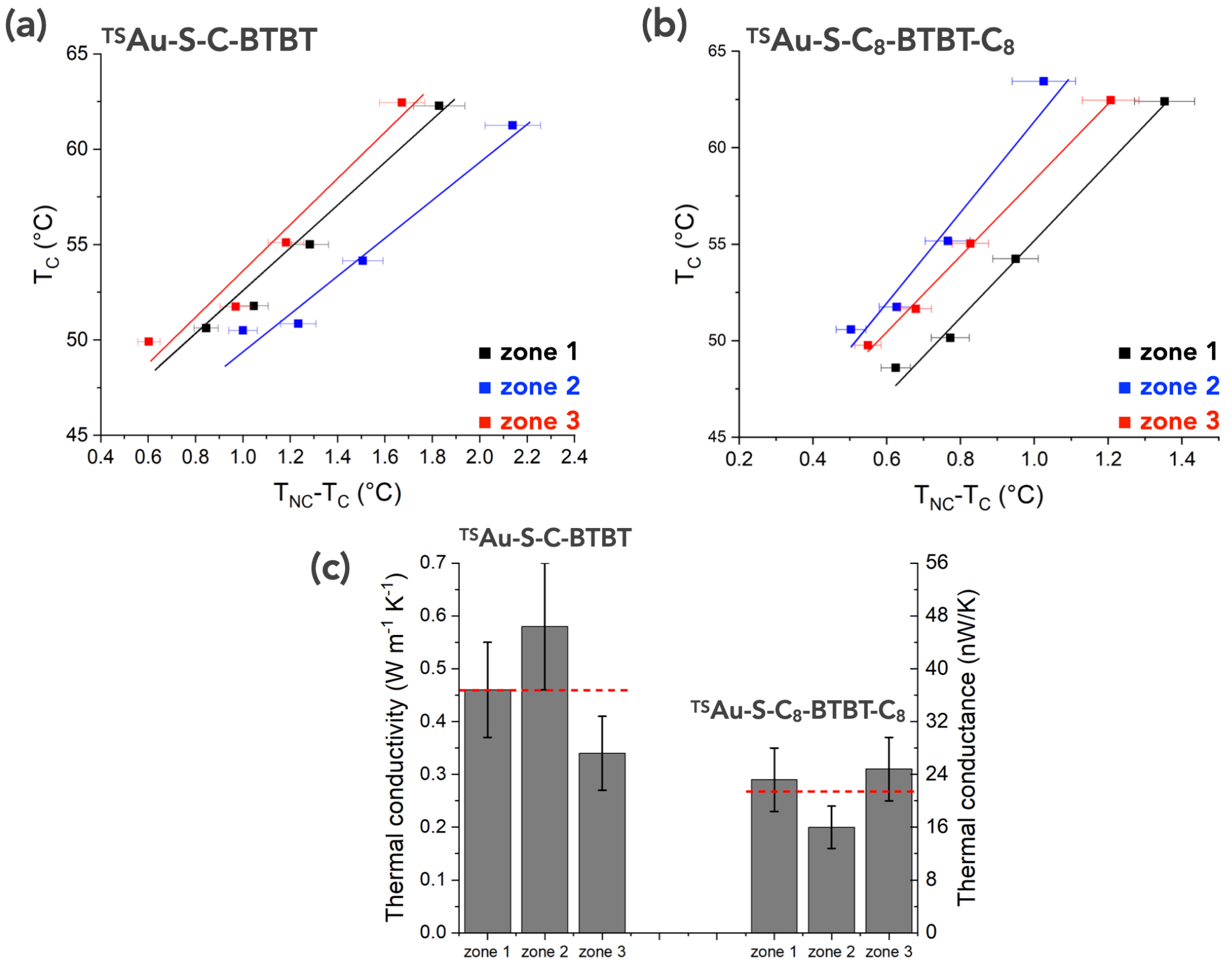

***Figure 3.*** *Tip temperature at contact, $T_C$, versus the temperature jump, $T_{NC}$ - $T_C$ measured at various heating of the tip ($T_C$ increases with the voltage applied on the Wheatstone bridge, $V_{WB}$=*

*0.7; 0.8, 0.9 and 1.1 V). The measurements were done at 3 locations (randomly chosen): **(a)** on the $^{TS}$Au-S-C-BTBT and **(b)** on the $^{TS}$Au-S-$C_8$-BTBT-$C_8$ SAMs. The lines are linear fits from which the thermal conductivity is determined (Eq. (1)). **(c)** Thermal conductivity, $\kappa_{SAM}$, and thermal conductance, $G_{th,SAM}$, for the three zones of the two MJs. The dashed red lines indicate the mean values (see Table 1).*

| | $^{TS}$Au-S-C-BTBT | | | $^{TS}$Au-S-$C_8$-BTBT-$C_8$ | | |
|---|---|---|---|---|---|---|
| | $\kappa_{SAM/Au}$ (W m$^{-1}$ K$^{-1}$) | $\kappa_{SAM}$ (W m$^{-1}$ K$^{-1}$) | $G_{th,SAM}$ (nW/K) | $\kappa_{SAM/Au}$ (W m$^{-1}$ K$^{-1}$) | $\kappa_{SAM}$ (W m$^{-1}$ K$^{-1}$) | $G_{th,SAM}$ (nW/K) |
| zone #1 | 7.67 | 0.46±0.08 | 37±6 | 1.76 | 0.29±0.06 | 23±5 |
| zone #2 | 9.75 | 0.58±0.12 | 46±10 | 1.23 | 0.20±0.04 | 16±3 |
| zone #3 | 3.71 | 0.34±0.07 | 27±5 | 1.82 | 0.31±0.06 | 25±5 |
| mean | | 0.46±0.27 | 37±21 | | 0.27±0.16 | 22±13 |

***Table 1.*** *Values of the measured thermal conductivity of the SAM/Au samples, $\kappa_{SAM/Au}$; thermal conductivity of the SAM removing the Au substrate contribution, $\kappa_{SAM}$, and the corresponding thermal conductance of the SAMs, $G_{th,SAM}$ (see text).*

The electron transport (ET) properties of the same SAMs were also measured by conductive-AFM (see details of the measurement protocols and data analysis in the Supporting Information). Figures 4-a and 4-b show the 2D histograms ("heat map") of the current-voltage (*I-V*) characteristics measured on the $^{TS}$Au-S-C-BTBT/PtIr C-AFM tip and $^{TS}$Au-S-$C_8$-BTBT-$C_8$/PtIr C-AFM tip molecular junctions (MJs). Due to the presence of the C8 alkyl chains at the α,ω positions of the BTBT, the conductance of the $C_8$-BTBT-$C_8$ MJs is much lower and no current is measurable in the applied voltage range from -1.5 V to 1.5 V (below the sensitivity limit of the C-AFM instrument ≲ 2-3 pA); this is due to the decrease in the electron transmission probability at the Fermi energy, $T(\varepsilon_F)$, by a factor ≈$10^4$ (see simulation section). For the BTBT MJs, the *I-V* dataset was analyzed with the single energy-level model (SEL model, Eq. S6 in the Supporting Information), which give the energy position $\varepsilon_0$ of the molecular orbital (here the highest occupied molecular orbital, HOMO) involved in the electron transport as well as the electronic coupling energies with the two electrodes $\Gamma_1$ and $\Gamma_2$ (hybridization between the molecular orbitals and the electron density of states in the electrodes). Figure 4-c shows the statistical distributions of the $\varepsilon_0$ parameter obtained by fitting the SEL model on all individual *I-V* traces of the dataset shown in Fig. 3-a. Note that such an analysis is not possible for the $C_8$-BTBT-$C_8$ MJs due to the lack of a measurable current in the -1.5 V/1.5 V window (Fig. 4-b). The $\varepsilon_0$ statistics show a normal distribution with a main value at $\varepsilon_0$ = 0.59 ± 0.04 eV, with a small tail at higher energy values. The zero-bias conductances of the SAMs (first derivative of the I-Vs in Fig. 4-a) show a very broad dispersion of the values. It can be fit by a log-normal distribution with a mean conductance values of $\langle G_{SAM} \rangle$=1.2x10$^{-9}$ S. A great dispersion is also observable from the statistical distribution of the current at a given voltage (see Fig. S4 in the Supporting Information), which are broad with a large tail at lower currents. These experimental behaviors of the conductance/current values result in a broadly distributed values of the electrode coupling energies $\Gamma_1$ and $\Gamma_2$ (0.1 to 10 meV, Fig. S5) with the SEL model fits. These features may indicate the existence of several molecular organizations in the SAM and/or configurations at the molecule/SAM interface (*vide infra*, simulations section).

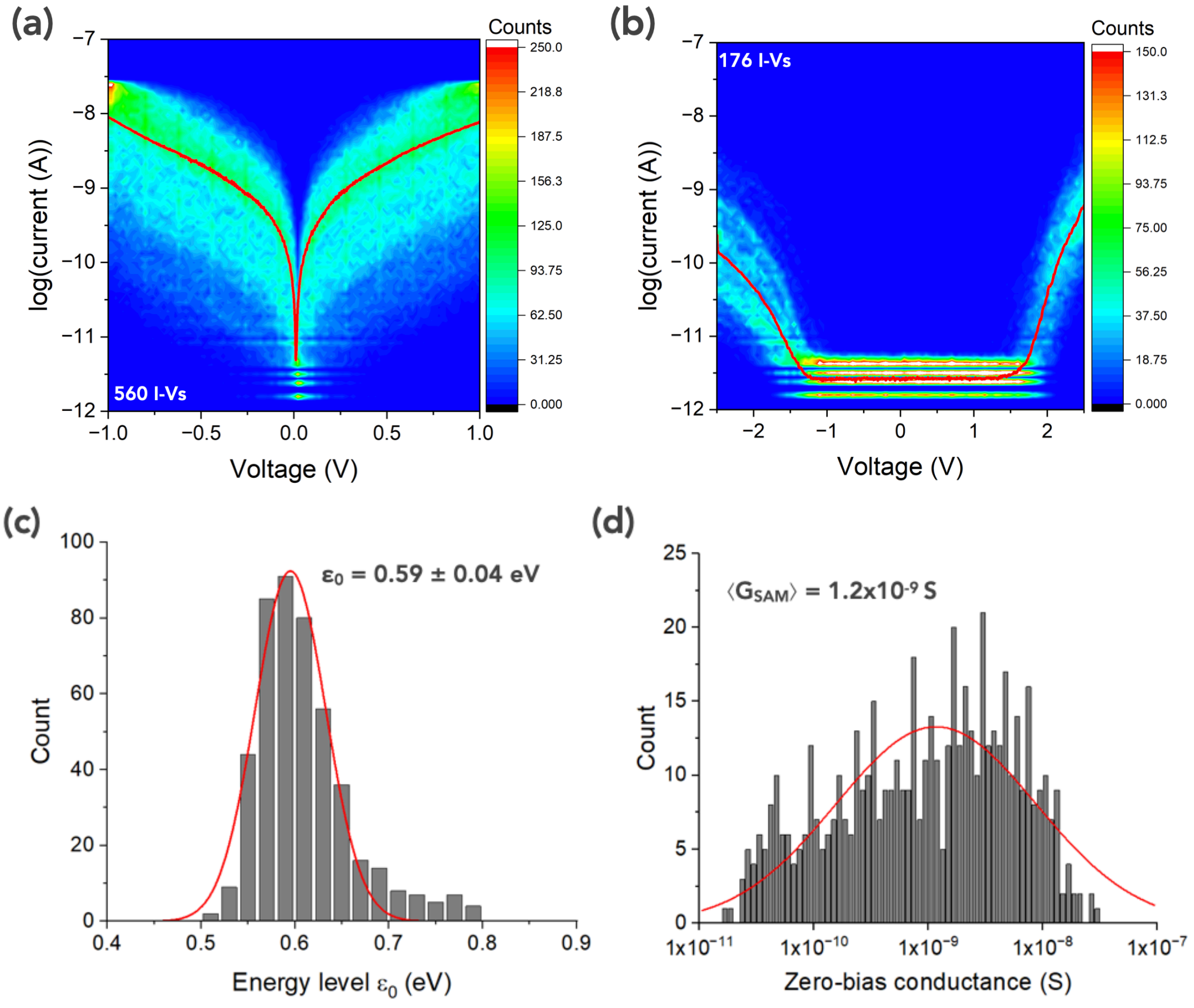

***Figure 4.*** *2D histograms of the I-V curves:* ***(a)*** *$^{TS}$Au-S-C-BTBT/PtIr C-AFM tip molecular junctions;* ***(b)*** *$^{TS}$Au-S-$C_8$-BTBT-$C_8$/PtIr C-AFM tip molecular junctions. The red lines are the mean $\bar{I}$-V curves.* ***(c)*** *Distribution of the energy values $\varepsilon_0$ for the dataset of the $^{TS}$Au-S-C-BTBT/PtIr C-AFM tip molecular junctions. The solid red lines are the fit with a normal distribution with the mean value ± standard deviation shown in the panel.* ***(d)*** *Statistical distribution of the zero-bias SAM conductance. The red line is a fit by a log-normal distribution with log-mean, log-μ=-8.93 (or mean $\langle G_{SAM} \rangle = 1.2\times10^{-9}$ S), log standard deviation, log-σ=0.87.*

In order to shed light on the experimental data, we calculated the energy dependent transmission probability, *T(ε)*, for the two MJs using DFT (density functional theory), see details in the Supporting Information. To explain the broad dispersion of ET in the $^{TS}$Au-S-C-BTBT MJ (Figs. 4-c and d), we simulated *T(ε)* for several conformations of the BTBT molecule in the SAMs. The structural model of the $^{TS}$Au-S-C-BTBT MJ is built by considering tilted molecules with a density of molecules deduced from the ellipsometry measurements (≈ 0.45 $nm^2$/molecule, see Tables S1). We first considered two conformations, with the molecules tilted with their edge toward the Au surface (referred to as "tilted1") or their aromatic plane toward the surface ("tilted2") - Fig. 5-a. Indeed, thiolated SAMs planar aromatic rings often organize as ordered domains differing by their organization.[30, 31] The calculated *T(ε)* shows a small energy difference for the HOMO with respect to the Fermi level (Fig. 5-b), the "tilted2" configuration gives a HOMO deeper in energy by 0.1 eV (HOMO at 0.4 eV for "tilted1" and 0.5 eV for "tilted2"). The calculated total energy of the "tilted1" configuration is lower than the one of the "tilted2" configuration by 0.31 eV. The "tilted1" configuration is the most stable and therefore the most probable configuration. We note that, albeit the calculated energies (0.4 and 0.5 eV, Fig. 5-b) do not exactly match the experimental values (0.59 eV, Fig. 4-c), the calculated energy shift between the two matches the 0.2-0.3 eV experimental dispersion. We note that the amplitude of *T(ε)* in the energy window -1/+1 eV is different for the two configurations that may contribute to the dispersion of the measured I-V curves assuming a mix of the two configurations in the SAM.

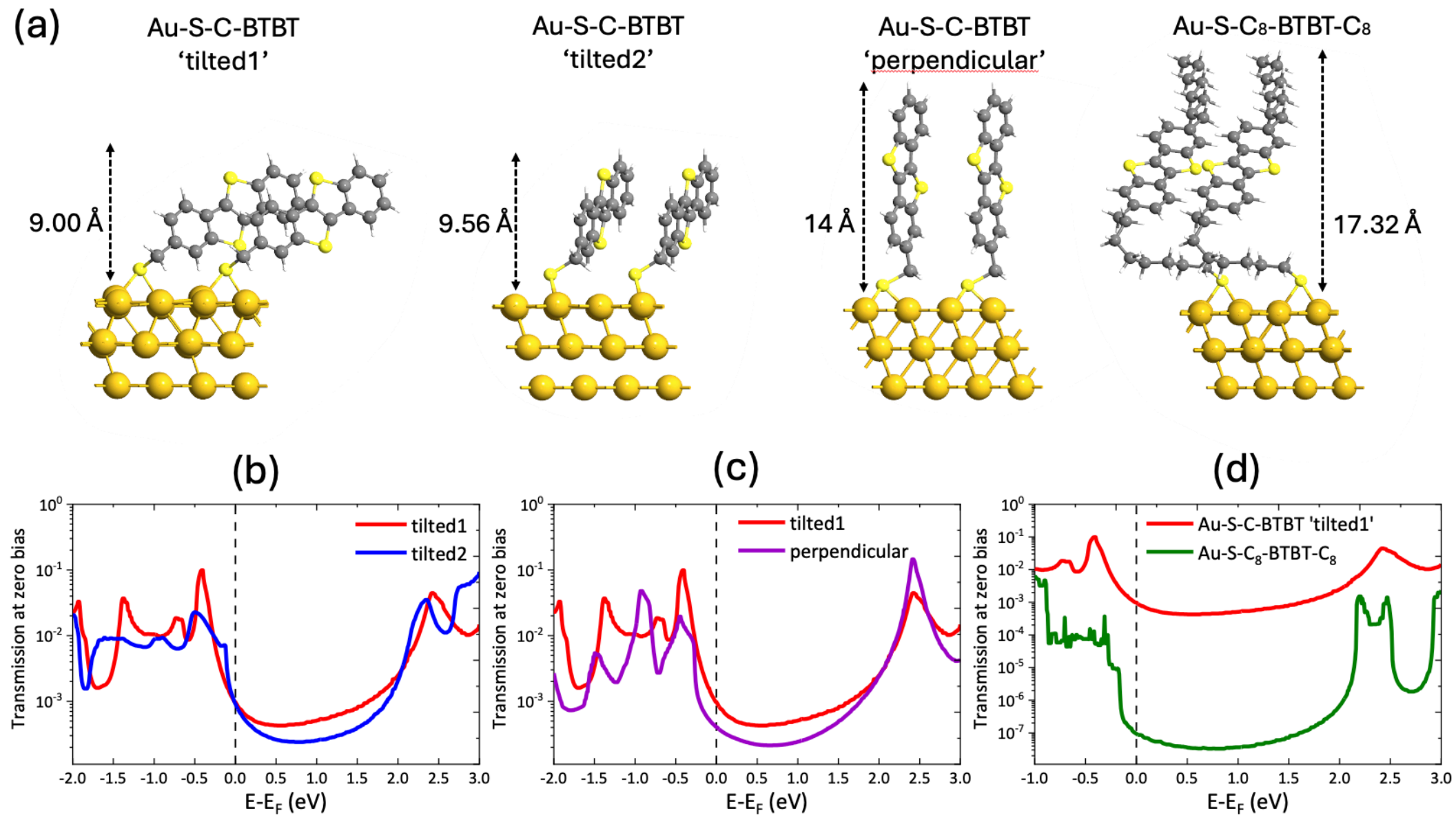

***Figure 5***. ***(a)*** *Schemes of the various structural models of the Au-molecule SAMs.* ***(b)*** *Calculated transmission coefficient at zero bias for the TSAu-S-C-BTBT SAM with two configurations for the tilted molecules: the molecule edge toward the Au surface (red line, "tilted1") or the face toward the surface (blue line, "tilted2").* ***(c)*** *Calculated transmission coefficient at zero bias for the TSAu-S-C-BTBT SAMs with the maximum packing density (see text) and the molecules perpendicular to the surface (purple line).* ***(d)*** *Comparison of the calculated transmission coefficient at zero bias for the TSAu-S-C-BTBT MJ ("tilted1" configuration, red line) and for the TSAu-S-$C_8$-BTBT-$C_8$ MJ (green line).*

We also considered another hypothesis. The SAM thickness measurement by ellipsometry gives an average value over a large area (light spot on the order of $mm^2$), while the C-AFM tip probes a tiny area (≈ 10 $nm^2$, see the Supporting Information). Therefore, if small clusters of more densely packed molecules are embedded in an overall less dense SAM, the C-AFM dataset can include I-V traces recorded on such denser clusters. The coexistence of nanoscale domains with different molecular organization is the consequence of the growth mechanisms of the SAMs.[30, 32] We calculated *T(ε)* for a SAM at its maximum packing density (≈0.26 $nm^2$/molecule, Table S1), with the molecules nearly perpendicular to the Au surface (Fig. 5-c). In that case, the HOMO is only very weakly shifted (at 0.45 eV), but another transmission peak (HOMO-1) with a large amplitude appears close to it (≈0.9 eV), *i.e.* readily accessible in the energy window used for the I-V measurements. In this case, when the I-V curves in the dataset coming from such compact packing clusters are analyzed with the SEL model, the fit likely returns a slightly higher "effective" energy $\varepsilon_0$ value, giving rises to the higher energy tail in the distribution (Fig. 4-c). The values of *T(ε)* in the HOMO-LUMO gap are also different, again contributing to the dispersion of the experimental conductance/current. For the $^{TS}$Au-S-$C_8$-BTBT-$C_8$ MJs, we calculated *T(ε)* considering the same density of molecules as for the $^{TS}$Au-S-C-BTBT MJs, as deduced from the thickness measurements (Table S1). The HOMO lies at about the same energy as for the $^{TS}$Au-S-C-BTBT MJs (Fig. 5-c) and as expected *T(ε)* at the Fermi energy is decreased by a factor ≈ $10^4$ in consistency with the decrease in the measured current (> $10^3$ in the -1 to 1 V window, Fig. 4). Finally, from the slope of *T(ε)* at the Fermi energy, we calculated the Seebeck coefficient as[1]

$$S = -\frac{\pi^2}{3}\frac{k^2 T}{e}\left(\frac{\partial lnT(\varepsilon)}{\partial \varepsilon}\right)_{\varepsilon_F} \quad (2)$$

with *k* the Boltzmann constant, *e* the electron charge, *T* the temperature. Doing so, we obtain $S \approx 36$ µV/K and 245 µV/K for the $^{TS}$Au-S-C-BTBT (from the calculated *T(ε)* of the "tilted1" configuration) and $^{TS}$Au-S-$C_8$-BTBT-$C_8$ MJs, respectively.

In a simple model, the difference between the thermal resistance of the $^{TS}$Au-S-$C_8$-BTBT-$C_8$ MJ and the $^{TS}$Au-S-C-BTBT MJ is twice the thermal resistance of the $C_8$ alkyl chain (see the Supporting Information, Fig. S6). Given the estimated number of molecules contacted by the SThM tip (~ 2500, see the Supporting Information), the thermal conductance per molecule deduced from the measurements of $G_{th,SAM}$ (Table 1) are $G_{th,mol} \approx 15$ pW/K for BTBT and ≈ 8.8 pW/K for $C_8$-BTBT-$C_8$, from which we infer a thermal conductance of ≈ 42 pW/K for the $C_8$ alkyl chain in our MJs. We note that this crude estimation is fairly of the same order of magnitude as the previously measured values for octane chains in SAM-based MJs (≈ 14 pW/K)[7] and single molecule experiments (≈ 26 pW/K and 37 pW/K).[8, 14] This simple estimation assumes that the molecule-electrode interface thermal resistance (interface Kapitza resistance)[33, 34] is the same at the covalent bottom interface (Au-S-C-) and at the top molecule/tip interface (-$CH_3$/Pd SThM tip or -phenyl/Pd SThM tip), which is not rigorously exact since the nature of the molecule-interface matters (covalent *vs.* van der Waals). It was shown for alkane-based MJs (using time-domain thermal reflectance) that van der Waals molecule-electrode interfaces result in lower thermal conductance than a covalent molecule-electrode interface (by a factor of ~ 2).[35] The chemical nature of the electrode metal matters too (*e.g.* the thermal conductance of Au-alkanedithiol-Au is *ca.* twice that of Au-alkanedithiol-Pd due to interface vibrational mismatch).[36] Nevertheless, this general agreement with previous measurements validate the present data. For completeness, the electronic contribution to the thermal conductance was estimated and, as expected,[37] is negligible. From the C-AFM measurements (Fig. 4), the mean electronic conductance per molecule at zero bias is $G_{e,mol} \approx 8\text{x}10^{-11}$ S ($1\text{x}10^{-6} G_0$, with $G_0$ the quantum of conductance $G_0 = 2e^2/h$ = 77.5 µS) for the $^{TS}$Au-S-C-BTBT (considering the mean conductance peak, $G_{e,SAM}$ = 4.1x10$^{-9}$ *S*, Fig. 4-c, with ~ 15 molecules in the C-AFM MJ, see the Supporting Information). Assuming that the Wiedemann-Franz law holds in MJs and at atomic scale[8, 14, 38] (which is still an open question at the nanoscale in general),[39, 40] we deduced an electronic contribution to the thermal conductance $G_{th,elec}=G_{e,mol}L_0T$ = 6x10$^{-4}$ pW/K, with $L_0$ the Lorenz number (2.44x10$^{-8}$ W Ω K$^{-2}$) and $T$ the temperature. Albeit there is uncertainty in the number of molecules in contact with both the C-AFM and SThM tips, this value of the electronic thermal conductance of BTBT is of the same order of magnitude as the calculated values for molecules like octanethiol and OPE3 (around 0.01 pW/K)[7] or as the smaller and negligible values for many molecules.[8, 41]

The thermal conductance of the single molecule (≈15 and ≈8.8 pW/K, for BTBT and $C_8$-BTBT-$C_8$, respectively) are on a par or smaller than that of alkyl chains and OPE: 15-30 pW/K (alkanedithiol, 2 to 10 carbon atoms),[8] 10-25 pW/K (alkanethiol, 2 to 18 carbon atoms),[7] ≈37 pW/K (8 carbon atoms),[14] ≈50 pW/K (2 to 24 carbon atoms)[4] and ≈23-24 pW/K for OPE derivatives.[11, 14] With a calculated Seebeck coefficient of 36 µV/K, a mean single molecule electrical conductance of 8x10$^{-11}$ S (*vide supra*), we estimate a mean $ZT \approx 2\text{x}10^{-6}$ (for BTBT at 300K). An upper limit can be estimated from the measured statistical distributions: with a maximal electron conduction $G_{e,mol}$ = 2.5x10$^{-9}$ S ($G_{SAM,max}$ = 3x10$^{-8}$ S, Fig. 4-d), and the lowest thermal conductance $G_{th,mol}$ ≈ 6.4 pW/K (minimal value of $G_{th,SAM}$ = 16 nW/K, Table 1), we can get $ZT_{max} \approx 1.1\text{x}10^{-4}$. These values are comparable with the complete thermoelectric characterization of OPE derivative MJs given $ZT$ (at 300K) ≈ 1.3-2x10$^{-5}$.[11] These values fall short compared to some theoretical, optimized, predictions (e.g. values from 3 to 4 have been theoretically predicted for Zn-porphyrin MJs)[42] and the required needs of $ZT > 1$ to envision practical applications.[10, 43]

The higher Seebeck coefficient of the $C_8$-BTBT-$C_8$ MJ is consistent with a decrease of the electrode coupling energies (due to the intercalation of the $C_8$ alkyl chains). Such a decrease in the electrode energy coupling is known to reduce the broadening of the molecular orbitals and thus to increase the slope of $T(\varepsilon)$ at $\varepsilon_F$. However, in the present case, this also induces a large decrease in the electron conductance (Fig. 4-b). Smaller alkyl chains (say, 3-4 carbon atoms) might be the compromise for not lowering the electronic conductance too much, while maintaining a thermal conductance below 10 pW/K and a Seebeck coefficient above 200 µV/K, these last two conditions being prerequisite to obtain high $ZT$ molecular devices.[10]

To sum up, we have combined experiment and theory to characterize the full thermoelectric properties of two benzothieno-benzothiophene (BTBT) SAM-based molecular junctions. We controlled the phononic thermal conductance by inserting alkyl chains (8 carbon atoms) between the BTBT core and the two electrodes. As previously demonstrated in thin films of the same molecules,[17] the alkyl chains efficiently reduced the phononic thermal transport and the thermal conductance of the $C_8$-BTBT-$C_8$ molecular junctions is decreased reaching a low value of 8.8 pW/K (per molecule vs. 15 pW/K in the BTBT molecular junction), one of the lowest thermal conductance for molecular junction so far measured.[41] Similarly, the efficient decoupling of the BTBT core from the electrodes leads to one of the highest expected Seebeck coefficient, $S$ = 245 µV/K, for a molecular device, [12, 13] but at the expense of a too drastic reduction of the electron conductance. Further molecular engineering and optimization are required to avoid this drawback (e.g. shorter alkyl spacers, other anchoring groups, functionalization with side groups).[43] The thermoelectric figure of merit (at 300 K) of the core BTBT molecular junction is determined in a range $ZT \approx 2\times10^{-6}$ to $10^{-4}$, on a par with the experimental values reported for the archetype molecular junction based on olygo(phenylene ethynylene) derivatives, $ZT \approx$ 1.3-$2\times10^{-5}$.[11]

## Associated content

**Supporting Information.**

The Supporting Information is available free of charge at https://pubs.acs.org/....

Details on the molecule synthesis, NMR and MS characterizations, fabrication of the samples (electrodes and SAMs), ellipsometry measurements, topographic AFM and roughness measurements, SThM measurement protocols and substrate correction of the thermal conductivity, C-AFM measurement procedures, data analysis, electron transport analytical model, DFT computational details.

## Author information

**Corresponding author.**

**Dominique VUILLAUME** - *Institute for Electronics Microelectronics and Nanotechnology (IEMN), CNRS, Av. Poincaré, Villeneuve d'Ascq, France.* Orcid: orcid.org/0000-0002-3362-1669; E-mail: dominique.vuillaume@iemn.fr

**Authors.**

**Sergio Gonzalez-Casal** - *Institute for Electronics Microelectronics and Nanotechnology (IEMN), CNRS, Av. Poincaré, Villeneuve d'Ascq, France.* Orcid: orcid.org/0000-0001-8559-2612; E-mail: sergiogc956@gmail.com

**Rémy Jouclas** - *Laboratory of Polymer Chemistry, Université Libre de Bruxelles, Bd. du triomphe, Bruxelles, Belgium.* Orcid: orcid.org/0000-0002-1346-6817; E-mail: remy.jouclas@ulb.ac.be

**Imane Arbouch** - *Laboratory for Chemistry of Novel Materials, U. Mons, Belgium.* Orcid: orcid.org/0000-0002-4593-0098; E-mail: imane.arbouch@umons.ac.be

**Yves Geerts** - *Laboratory of Polymer Chemistry, Université Libre de Bruxelles, Bd. du triomphe, Bruxelles, Belgium; and International Solvay Institutes of Physics and Chemistry, Université Libre de Bruxelles, Bd. du Triomphe, Bruxelles, Belgium.* Orcid: orcid.org/0000-0002-2660-5767; E-mail: yves.geerts@ulb.be

***Colin van Dyck*** - *Theoretical Chemical Physics group, University of Mons, Mons, Belgium.* Orcid: orcid.org/0000-0003-2853-3821; E-mail: colin.vandyck@umons.ac.be

***Jérôme Cornil*** - *Laboratory for Chemistry of Novel Materials, U. Mons, Belgium*. Orcid: orcid.org/0000-0002-5479-4227; E-mail: jerome.cornil@umons.ac.be

***Author Contributions***

S.G.C. carried out the SThM and C-AFM measurements. R.J. and Y.G. synthesized the molecules, and S.G.C. fabricated the SAMs. I.A., C.v.D. and J.C. performed the theoretical computations. S.G.C. and D.V. analyzed the SThM and C-AFM data. D.V. supervised the project. The manuscript was written by D.V. with the contributions and comments of all the authors. All authors have given approval of the final version of the manuscript.

***Note***

The authors declare no competing financial interest.

## Acknowledgements.

S.G.C and D.V. acknowledge support from the ANR (# ANR-21-CE30-0065, project HotElo). Y. G. is thankful to the Belgian National Fund for Scientific Research (FNRS) for financial support through research projects Pi-Fast (No T.0072.18), Pi-Chir (No T.0094.22), 2D to 3D (No 30489208), and CHISUB (No 40007495). Financial supports from the Fédération Wallonie-Bruxelles (ARC No 20061) is also acknowledged. The research in Mons is supported by the Belgian National Fund for Scientific Research (FRS-FNRS) via the EOS CHISUB project (No 40007495) and within the Consortium des Équipements de Calcul Intensif – CÉCI (grant number U.G.018.18), and by the Walloon Region (LUCIA Tier-1 supercomputer; grant number 1910247). J.C. is an FNRS research director.

## References.

# Thermoelectric Properties of Benzothieno-Benzothiophene Self-Assembled Monolayer in Molecular Junctions.

Sergio Gonzalez-Casal,[1] Rémy Jouclas,[2] Imane Arbouch,[3] Yves Geerts,[2,4] Colin van Dyck,[5] Jérôme Cornil,[3] and Dominique Vuillaume.[1*]

1) *Institute for Electronics Microelectronics and Nanotechnology (IEMN), CNRS, Av. Poincaré, Villeneuve d'Ascq, France.*

2) *Laboratory of Polymer Chemistry, Université Libre de Bruxelles, Bd. du triomphe, Bruxelles, Belgium*

3) *Laboratory for Chemistry of Novel Materials, U. Mons, Belgium.*

4) *International Solvay Institutes of Physics and Chemistry, Université Libre de Bruxelles, Bd. du Triomphe, Bruxelles, Belgium.*

5) *Theoretical Chemical Physics group, University of Mons, Mons, Belgium.*

** Corresponding author: dominique.vuillaume@iemn.fr*

## Supporting Information.

## I. Synthesis.

***General.*** All reagents were purchased from Sigma-Aldrich (now Merck), VWR, Acros, Alfa Aesar, TCI and Fluorochem and were used without further purification. Technical grade solvents were purchased from Chem-Lab and used as supplied. Anhydrous solvents as chloroform, dichloromethane, N,Ndimethylformamide and tetrahydrofurane were distilled using common methods. Air- and/or moisture-sensitive liquids and solutions were transferred via a syringe or a Teflon cannula. Analytical thin-layer chromatography (TLC) was performed on aluminum plates with 10-12 µm silica gel containing a fluorescent indicator (Merck silica gel 60 F254). TLC plates were visualized by exposure to ultraviolet light (254 nm and 365 nm). Flash column chromatography was performed on Grace Davisil LC60A (70-200µm) silica. All NMR spectra were recorded on Jeol 400 MHz spectrometer. Chemical shifts are reported in parts per million (ppm, δ scale) from tetramethylsilane for 1H NMR (δ 0 ppm in chloroform and 1,1,2,2-tetrachoroethane) and from the solvent carbon for 13C NMR (e.g., δ 77.16 ppm for chloroform). The data are presented in the following format: chemical shift, multiplicity (s = singlet, d = doublet, t = triplet, m = multiplet), coupling constant in hertz (Hz), signal area integration in natural numbers, assignment (italic). The Mass Spectrometry analyses have been performed in the Organic Synthesis & Mass Spectrometry Laboratory at the University of Mons (Prof. Pascal Gerbaux) using MALDI-MS on a Q-TOF Premier mass spectrometer (Waters, Manchester, UK) in the positive ion mode.

### *1. Synthesis of BTBT-2-methanethiol (HS-C-BTBT)*

**2-formyl-BTBT 2** [1]

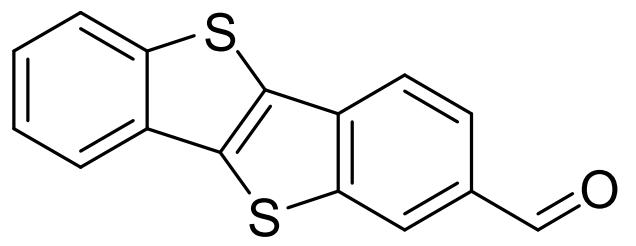

Chemical Formula: $C_{15}H_8OS_2$
Exact Mass: 268,0017
Molecular Weight: 268,3480

To a suspension of BTBT **1** (3.00 g, 12.5 mmol, 1 eq.) in dry DCM (250 mL) at -64°C was added aluminum chloride (7.49 g, 56.2 mmol, 4.5 eq.) and dichloromethyl methyl ether (1.58 mL, 17.8 mmol, 1.4 eq.) and the mixture was stirred at this temperature for 6h. Then the reaction was quenched by addition of water (200 mL), and the mixture was neutralized by addition of aqueous saturated $Na_2CO_3$. The phases were separated, and the aqueous layer was extracted with $CHCl_3$ (300 mL), the combined organic layers were dried over $MgSO_4$ and evaporated under reduced pressure. $^1$H NMR of the crude mixture enabled us to assess a 2/4 regioselectivity of formylation of ~ 7. The resulting crude product was adsorbed on silica and eluted with toluene on a silica pad, and the resulting mixture was subjected to column chromatography (Petroleum ether / DCM 9:1 to 1:1) to afford 821 mg of 2-formyl-BTBT **2** with 24% yield, and 344 mg of unreacted BTBT. Comparison of the NMR spectrum of the crude compound with the target product suggests that a substantial part of it has been lost during the purification by column chromatography, thus increasing the gradient to 100% DCM should probably enable to get the remaining aldehyde.

**$^{1}$H NMR (400 MHz, CHLOROFORM-*D*) δ** : 10.13 (s, 1H), 8.46 (s, 1H), 8.02-7.95 (m, 4H), 7.53-7.46 (m, 2H).

**<u>2-methanol-BTBT 3</u>**

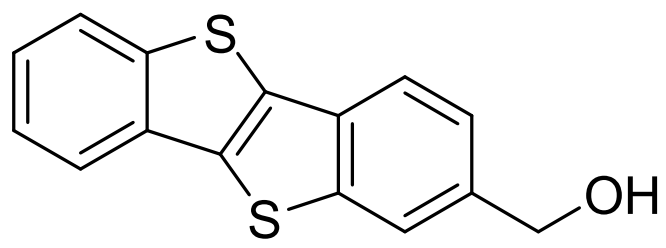

Chemical Formula: $C_{15}H_{10}OS_2$
Exact Mass: 270,0173
Molecular Weight: 270,3640

To a solution of 2-formyl-BTBT **2** (821 mg, 3.05 mmol, 1 eq.), in THF/MeOH 2:1 (300 mL) at 0°C was added sodium tetraborohydride (255 mg, 6.73 mmol, 2.2 eq.) portionwise, and the mixture was stirred at this temperature for 30 min and at room temperature for 30 min more. Then the mixture was diluted with water (100 mL), then with 1M HCl (67 mL), the volatiles were evaporated and the resulting aqueous layer was extracted with chloroform (330 mL then 2 x 130 mL). The combined organic layers were dried over $MgSO_4$, the solvent was removed under reduced pressure and the resulting crude product was subjected to column chromatography (Toluene/EtOH 99:1 to 95:5) to afford 735 mg of pure product with 89% yield.

**$^{1}$H NMR (400 MHz, CHLOROFORM-*D*) δ** : 7.95 (s, 1H), 7.93 (d, *J* = 7.9 Hz, 1H), 7.89 (d, *J* = 7.5 Hz, 1H), 7.88 (d, *J* = 8.2 Hz, 1H), 7.49-7.39 (m, 3H), 4.87 (d, *J* = 5.9 Hz, 2H), 1.75 (t, *J* = 5.9 Hz, 1H).

**$^{13}$C NMR (101 MHz, CHLOROFORM-*D*) δ** : 142.78, 142.39, 138.19, 133.81, 133.36, 133.26, 132.72, 125.20, 125.08, 124.32, 124.20, 122.49, 121.80, 121.74, 65.50.

**HRMS (EI-GCMS)** : m/z = 270.0169. calcd for $C_{15}H_{10}OS_2$ : 270.0173 $[M]^+$

**TLC (Toluene/EtOH 99:1)** : Rf = 12%.

**BTBT-2-methanethiouronium chloride 4**

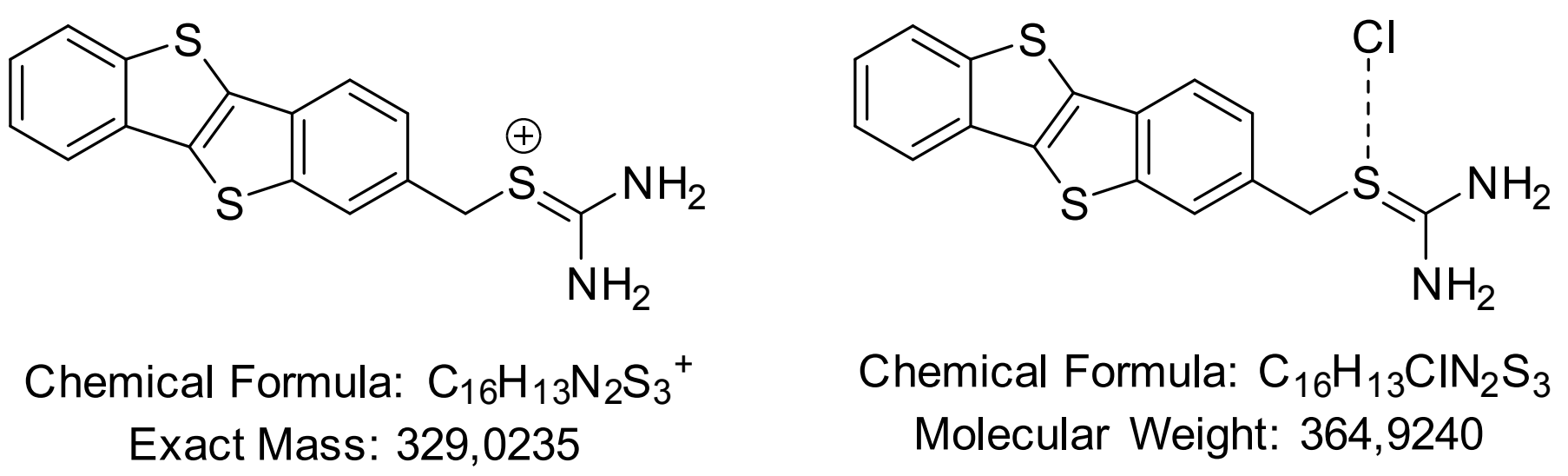

To a suspension of 2-methanol BTBT **3** (607 mg, 2.25 mmol, 1 eq.) in *i*PrOH/HCl 6M 5:1 (240 mL) was added thiourea (683 mg, 8.98 mmol, 4 eq.) and the mixture was stirred at 100°C for 6 days. Then the mixture was cooled to room temperature, the precipitate was filtered, washed with *i*PrOH and dried over vacuum to afford 722 mg of pure BTBT-2-methanethiouronium chloride **4** as a white solid with 88% yield.

**$^1$H NMR (400 MHz, $(CD_3)_2SO$) δ** : 9.31 (d, *J* = 45.7 Hz, 4H), 8.23 (d, *J* = 1.5 Hz, 1H), 8.18-8.15 (m, 1H), 8.07 (d, *J* = 8.2 Hz, 1H), 8.07-8.05 (m, 1H), 7.59 (dd, *J* = 8.2, 1.6 Hz, 1H) 7.56-7.48 (m, 2H), 4.72 (s, 2H).

**$^{13}$C NMR (101 MHz, $(CD_3)_2SO$) δ** : 168.91, 141.79, 141.67, 133.39, 132.74, 132.61, 132.23, 131.93, 126.48, 125.73, 125.47, 124.79, 124.56, 122.03, 121.78, 34.33.

**HRMS (ESI-GCMS)** : m/z = 329.048. calcd for $C_{16}H_{13}N_2S_3$ : 329.0241 $[M]^+$

**M.p :** 243 ± 1°C (degradation).

**BTBT-2-methanethiol (HS-C-BTBT)**

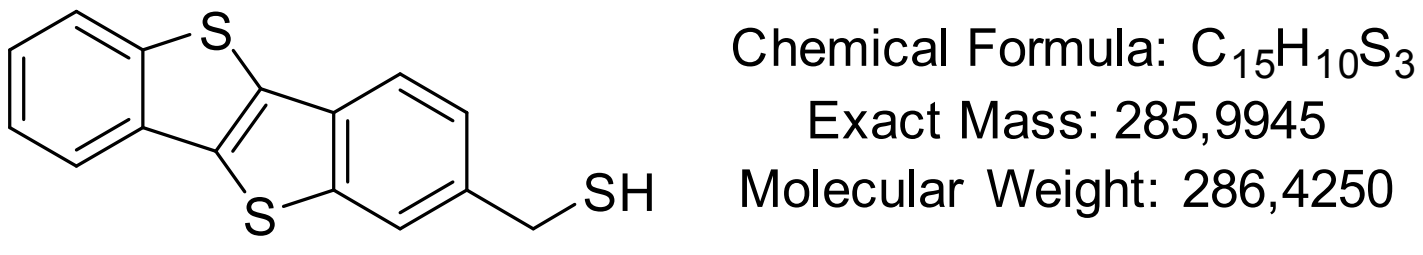

To a suspension of BTBT-2-methanethiouronium chloride **4** (48.5 mg, 133 µmol, 1 eq.) in *i*PrOH (15 mL) was added 1M aqueous NaOH (140 µL, 140 µmol, 1.05 eq) and the mixture was refluxed for 30 min. After cooling to room temperature, the mixture was diluted with 0.1 M aqueous HCl (15 mL), and the *i*PrOH was evaporated under reduced pressure. The resulting aqueous mixture was extracted 3 times with $CHCl_3$, and the combined organic phases were dried over $MgSO_4$, and evaporated under reduced pressure to afford 24 mg of pure HS-C-BTBT with 63% yield.

**$^{1}$H NMR (400 MHz, $CDCl_3$) δ** : 7.92 (d, *J* = 8.0 Hz, 1H), 7.88 (d, *J* = 7.0 Hz, 1H), 7.88 (s, 1H), 7.84 (d, *J* = 8.2 Hz, 1H), 7.48-7.39 (m, 3H), 3.92 (d, *J* = 5.6 Hz, 2H), 1.86 (t, *J* = 5.6 Hz, 1H).

**$^{13}$C NMR (101 MHz, $CDCl_3$) δ** : 142.80, 142.38, 138.44, 133.71, 133.34, 133.24, 132.21, 125.51, 125.18, 125.06, 124.19, 123.38, 121.89, 121.71, 29.35.

**HRMS (EI-GCMS)** : m/z = 285.9955. calcd for $C_{15}H_{10}S_3$ : 285.9945 $[M]^+$

**TLC (Heptane/$CHCl_3$ 1:1)** : Rf = 57%.

## *2. Synthesis of 2-octyl-6-(η-thiooctyl)-BTBT (HS-C$_8$-BTBT-C$_8$)*

**2-octyl-6-(η-bromooctanoyl)-BTBT 6**

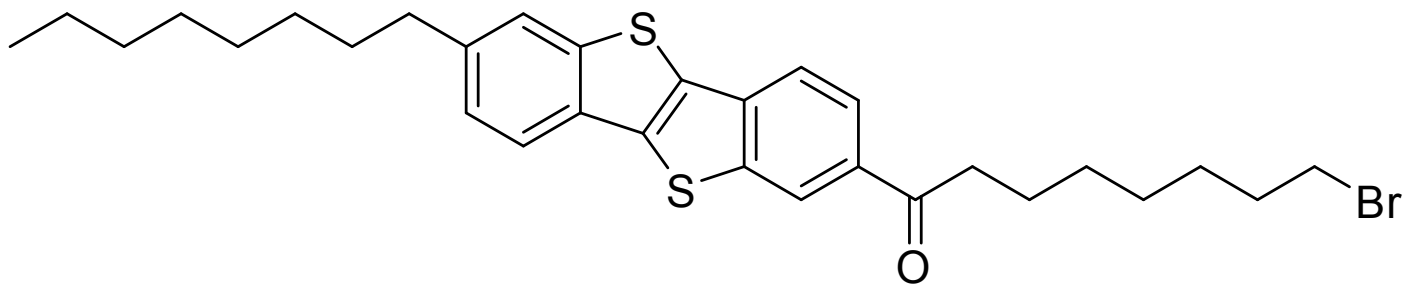

Chemical Formula: $C_{30}H_{37}BrOS_2$
Exact Mass: 556,1469
Molecular Weight: 557,6490

To a solution of η-bromooctanoic acid (791 mg, 3.55 mmol, 2.5 eq.) in dry DCM (35 mL) under argon atmosphere was added oxalyl chloride (426 µL, 4.96 mmol, 3.5 eq.) and 1 drop of DMF and the mixture was stirred at room temperature until no more bubbles could be seen (about 40 min). The volatiles were evaporated under reduced pressure to afford η-bromooctanoyl chloride that was dissolved in 20 mL of dry DCM and kept under argon atmosphere.

To a solution of 2-octyl-BTBT[6] **5** (500 mg, 1.42 mmol, 2 eq.) in dry DCM (15 mL) at -15°C was added aluminum chloride (520 mg, 3.90 mmol, 2.75 eq.) under argon atmosphere, then the mixture was cooled to -78°C and the solution of η-bromooctanoyl chloride was added dropwise over 30 min, and the mixture was allowed to reach room temperature under stirring overnight. The reaction was quenched by successive addition of ice and water (70 mL), the DCM was removed under reduced pressure, and the precipitate was filtered, washed with water and methanol. The resulting solid was washed with petroleum ether on a pad of silica, then eluted with a mixture of petroleum ether and chloroform 1:1 to afford 610 mg of pure 2-octyl-6-(η-bromooctanoyl)-BTBT **6** as a white solid with 77% yield.

**$^{1}$H NMR (400 MHz, CHLOROFORM-*D*) δ** : 8.54 (dd, *J* = 1.5, 0.6 Hz, 1H), 8.05 (dd, *J* = 8.3, 1.5 Hz, 1H), 7.90 (dd, *J* = 8.3, 0.7 Hz, 1H), 7.83 (d, *J* = 8.3 Hz, 1H), 7.74 (dd, *J* = 1.5, 0.8 Hz, 1H), 7.31 (dd, *J* = 8.2, 1.5 Hz, 1H), 3.42 (t, *J* = 6.8 Hz, 2H), 3.07 (t, *J* = 7.3 Hz, 2H), 2.78 (t, *J* = 7.8 Hz, 2H), 1.91-1.77 (m, 4H), 1.74-1.67 (m, 2H), 1.51-1.22 (m, 16H), 0.88 (t, *J* = 7.2, 6.7 Hz, 1H).

**$^{13}$C NMR (101 MHz, CHLOROFORM-*D*) δ** : 199.51, 143.33, 142.20, 141.55, 137.16, 136.62, 133.45, 132.30, 130.82, 126.34, 124.79, 124.64, 123.57, 121.87, 121.30, 38.78, 36.32, 34.07, 32.89, 32.02, 31.76, 29.61, 29.45, 29.39, 29.32, 28.79, 28.17, 24.50, 22.80, 14.24.

**HRMS (MALDI) :** m/z = 556.1483. calcd for $C_{30}H_{37}BrOS_2$ : 556.1469 $[M]^+$

**TLC (Hetpane/$CHCl_3$ 1:1)** : Rf = 19%

**<u>2-octyl-6-(η-bromooctyl)-BTBT 7</u>**

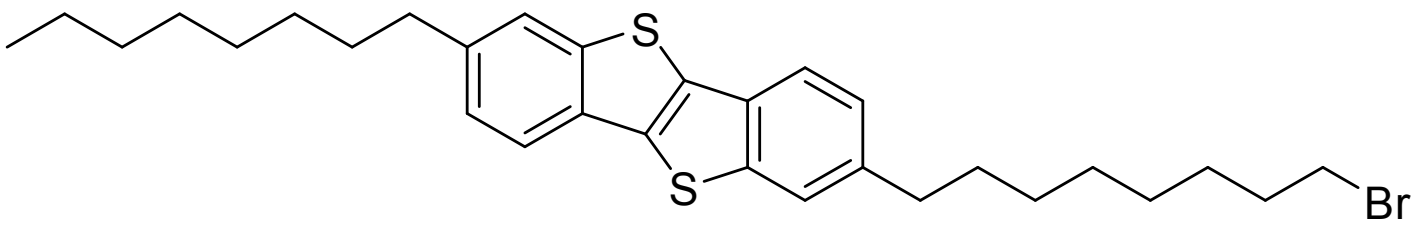

Chemical Formula: $C_{30}H_{39}BrS_2$
Exact Mass: 542,1677
Molecular Weight: 543,6660

To a suspension of $AlCl_3$ (103 mg, 771 µmol, 2.5 eq.) in anhydrous $Et_2O$ (1 mL) was added $LiAlH_4$ 1M/$Et_2O$ (1.03 mL, 771 µmol, 2.5 eq.) dropwise, then 2-octyl-6-(η-bromooctanoyl)-BTBT **6** (172 mg, 308 µmol, 1 eq.) in DCM (6 mL) was added dropwise and the mixture was stirred at room temperature overnight. The reaction was quenched by addition of ice, the phases were separated and the aqueous layer was extracted twice with DCM. The combined organic layers were dried over $MgSO_4$ and evaporated under reduced pressure to afford pure 2-octyl-6-(η-bromooctyl)-BTBT **7** (126 mg) as a white solid with 75% yield.

**$^{1}$H NMR (600 MHz, CHLOROFORM-*D*) δ** 7.76 (d, *J* = 8.1 Hz, 2H), 7.70 (m, 2H), 7.27 (ddd, *J* = 8.1, 3.4, 1.7 Hz, 2H), 3.40 (t, *J* = 6.9 Hz, 2H), 2.75 (t, *J* = 7.8 Hz, 4H), 1.88-1.83 (m, 2H), 1.72-1.67 (m, 4H), 1.45-1.25 (m, 18H), 0.90 (t, *J* = 7.0 Hz, 1H).

**$^{13}$C NMR (151 MHz, CHLOROFORM-*D*) δ** : 142.56 (2C), 140.23, 140.03, 132.71, 132.67, 131.37, 131.32, 125.97, 125.94, 123.46 (2C), 121.22, 121.20, 36.27, 36.21, 34.12, 32.94, 32.03, 31.85, 31.75, 29.63, 29.41, 29.29, 28.82, 28.29, 22.81, 14.25.

**HRMS (MALDI) :** m/z = 542.1666. calcd for $C_{30}H_{39}BrS_2$ : 542.1677 $[M]^+$.

**TLC (Petroleum ether)** : Rf = 24%.

**2-octyl-6-(η-thiooctyl)-BTBT (HS-$C_8$-BTBT-$C_8$)**

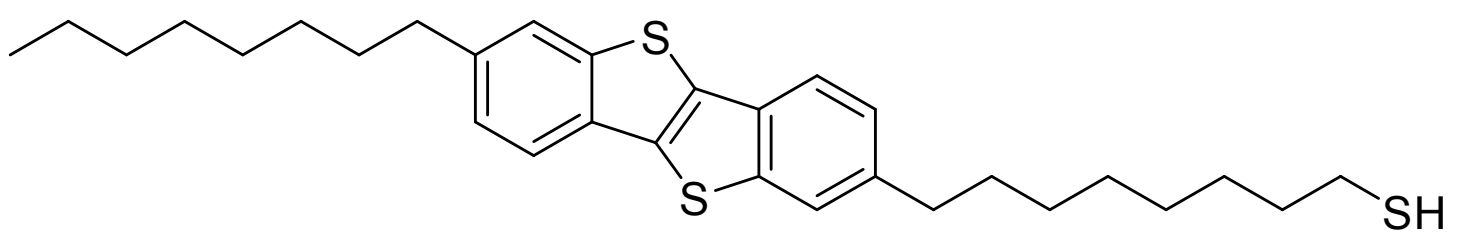

Chemical Formula: $C_{30}H_{40}S_3$
Exact Mass: 496,2292
Molecular Weight: 496,8300

To a suspension of 2-octyl-(η-bromooctyl)-BTBT **7** (192 mg, 353 µmol, 1 eq.) in *i*PrOH (35 mL) was added thiourea (80 mg, 1.06 mmol, 3.0 eq.) and the mixture was refluxed overnight. Then 1M aqueous sodium hydroxide (707 µL, 707 µmol, 2.0 eq.) was added and the reaction mixture was refluxed for 2h. Then the volatiles were removed under reduced pressure, the resulting mixture was diluted with water (20 mL) and extracted with chloroform (2x 20 mL). The combined organic layers were dried over $MgSO_4$, concentrated under vacuum and the resulting crude product was subjected to column chromatography (heptane/chloroform 9:1) to afford 95 mg of pure 2-octyl-6-(η-thiooctyl)-BTBT **(HS-$C_8$-BTBT-$C_8$)** with 54% yield.

**$^1$H NMR (400 MHz, CHLOROFORM-*D*) δ** : 7.76 (d, *J* = 8.1 Hz, 2H), 7.70 (s, 2H), 7.27 (dt, *J* = 8.1, 1.7 Hz, 2H), 2.75 (t, *J* = 7.9, 7.5 Hz, 4H), 2.52 (q, *J* = 7.5 Hz, 2H), 1.74-1.66 (m, 4H), 1.64-1.57 (m, 2H), 1.42-1.23 (m, 18H), 1.33 (t, *J* = 7.7 Hz, 1H), 0.90 (t, *J* = 7.0 Hz, 3H).

**$^{13}$C NMR (101 MHz, CHLOROFORM-*D*) δ** : 142.55 (2C), 140.22, 140.07, 132.70, 132.67, 131.36, 131.32, 125.96, 125.94, 123.45 (2C), 121.20, 36.26, 36.22, 34.15, 32.03, 31.85, 31.78, 29.63, 29.51, 29.47, 29.41, 29.33, 29.13, 28.49, 24.77, 22.81, 14.25.

**HRMS (MALDI) :** m/z = 496.2273. calcd for $C_{30}H_{40}S_3$ : 496.2292 $[M]^+$.

**TLC (Petroleum ether/$CHCl_3$ 9:1)** : Rf = 30%.

### 3. [1]H and [13]C NMR spectra

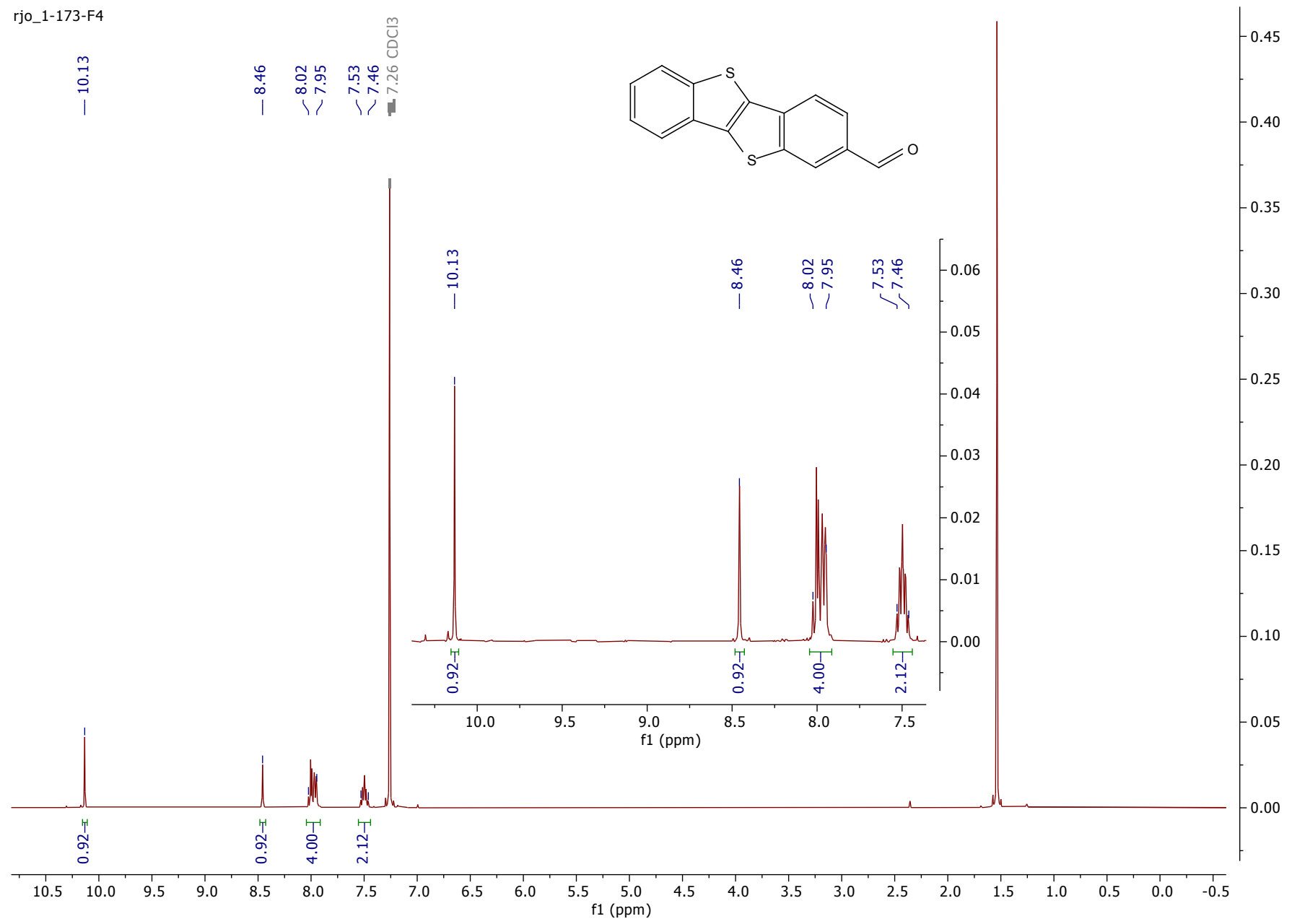

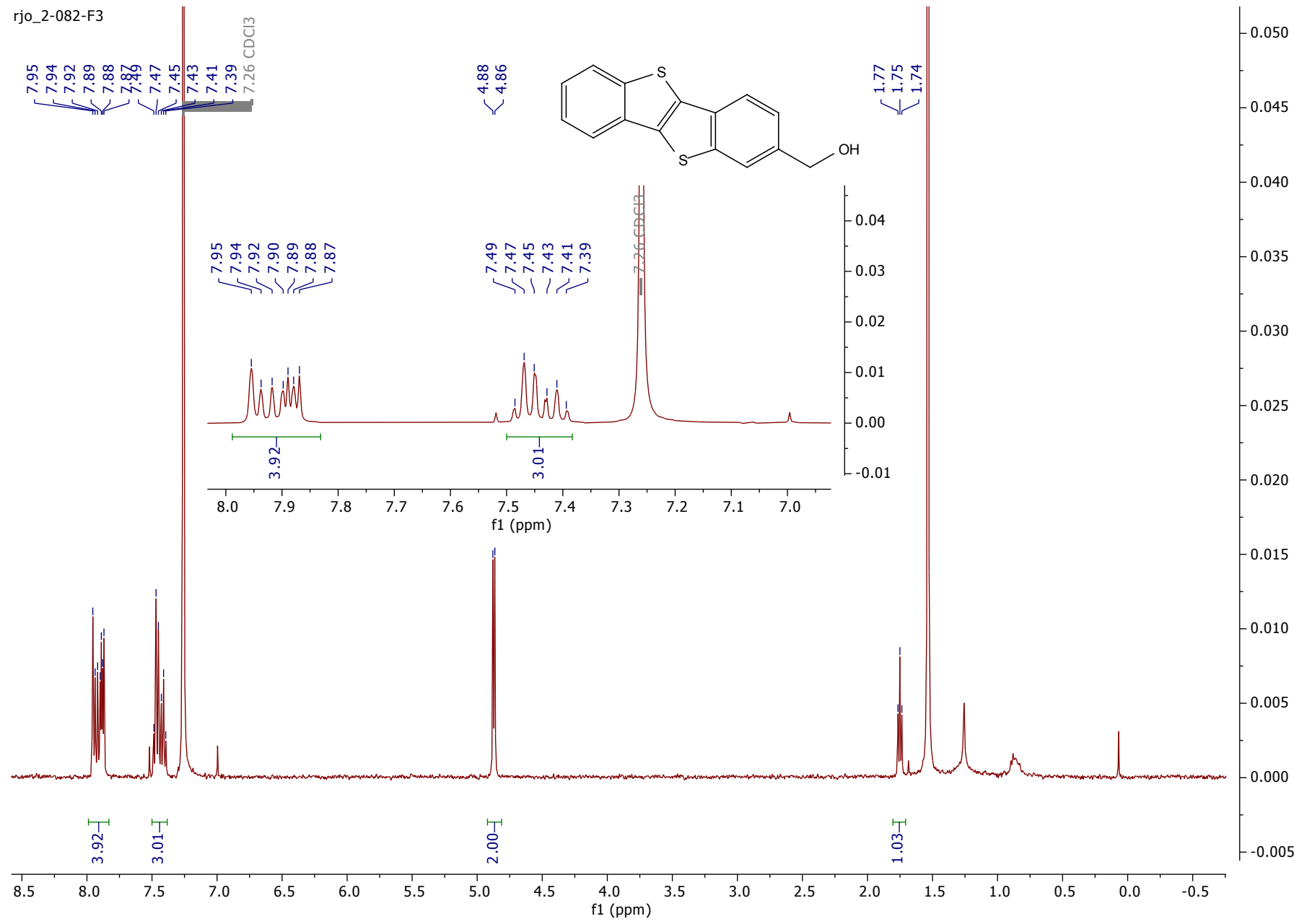

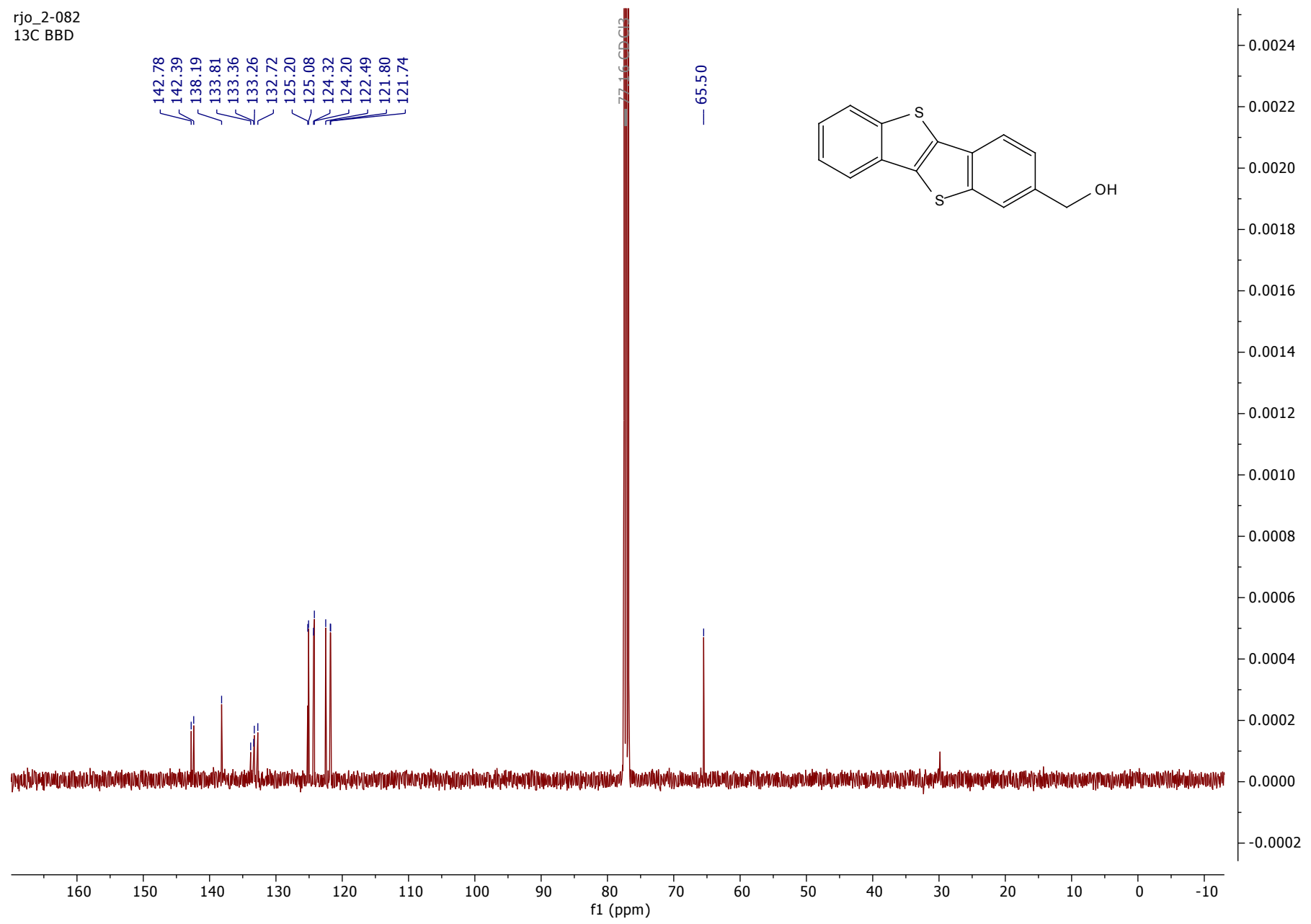

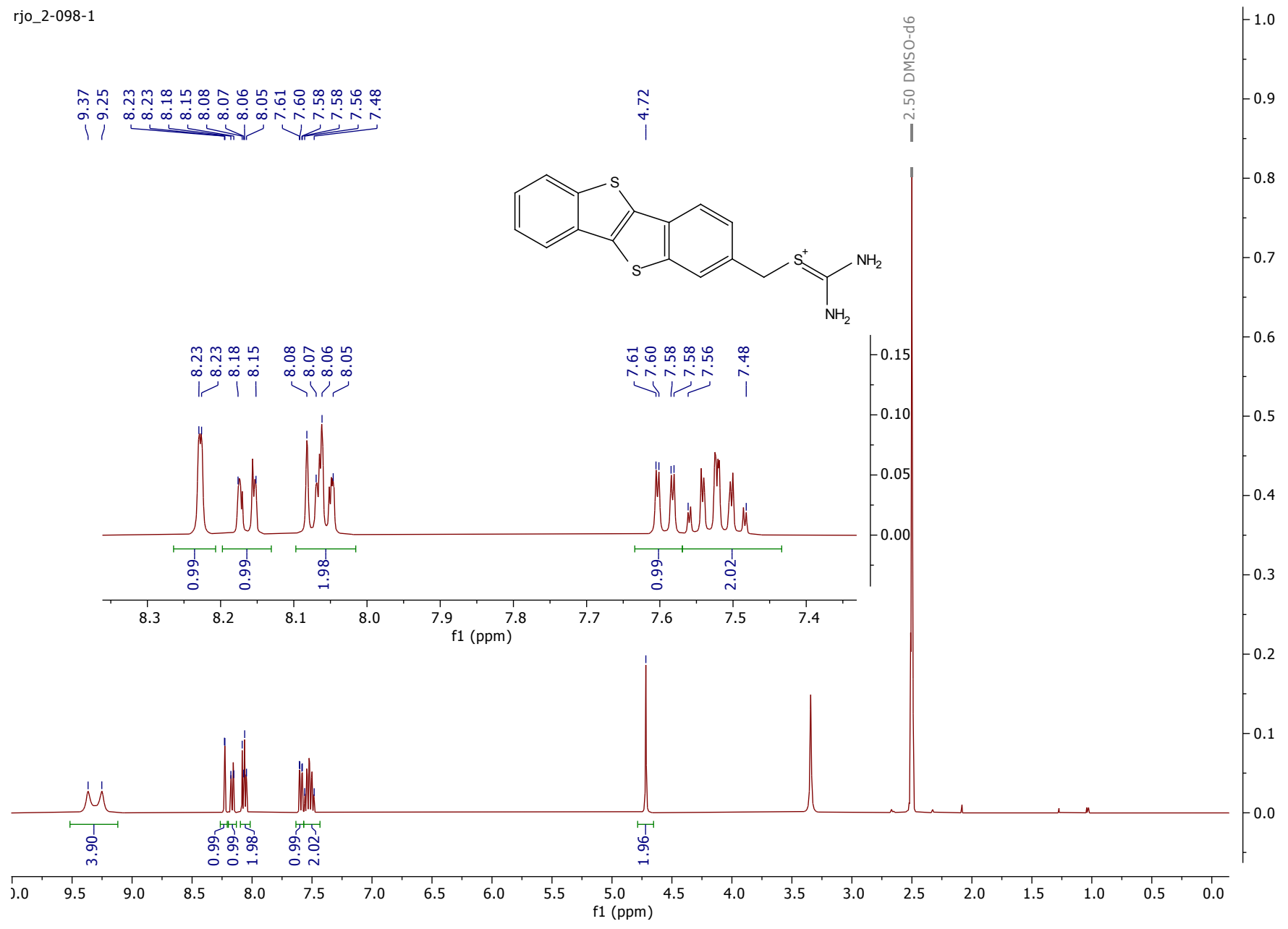

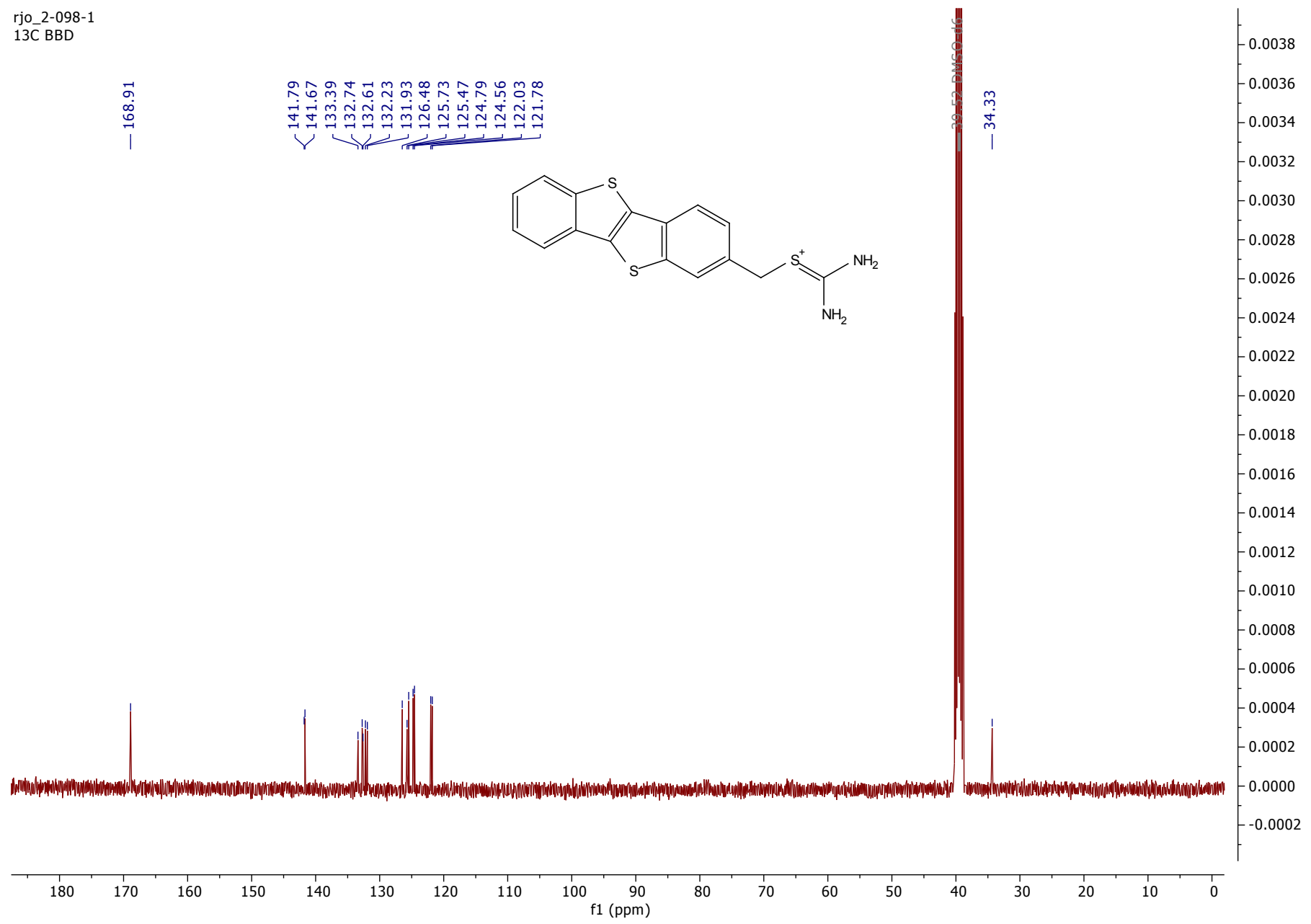

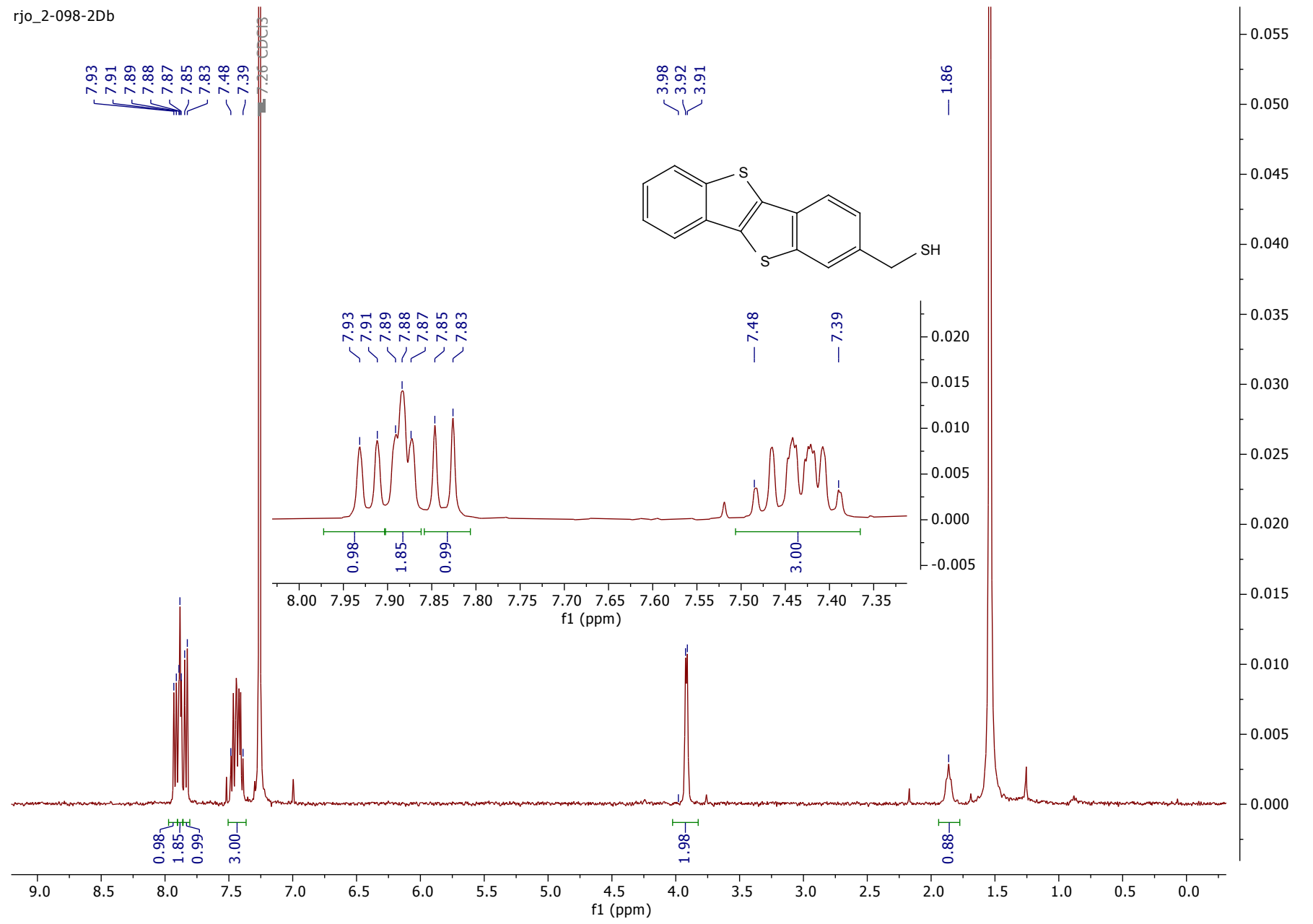

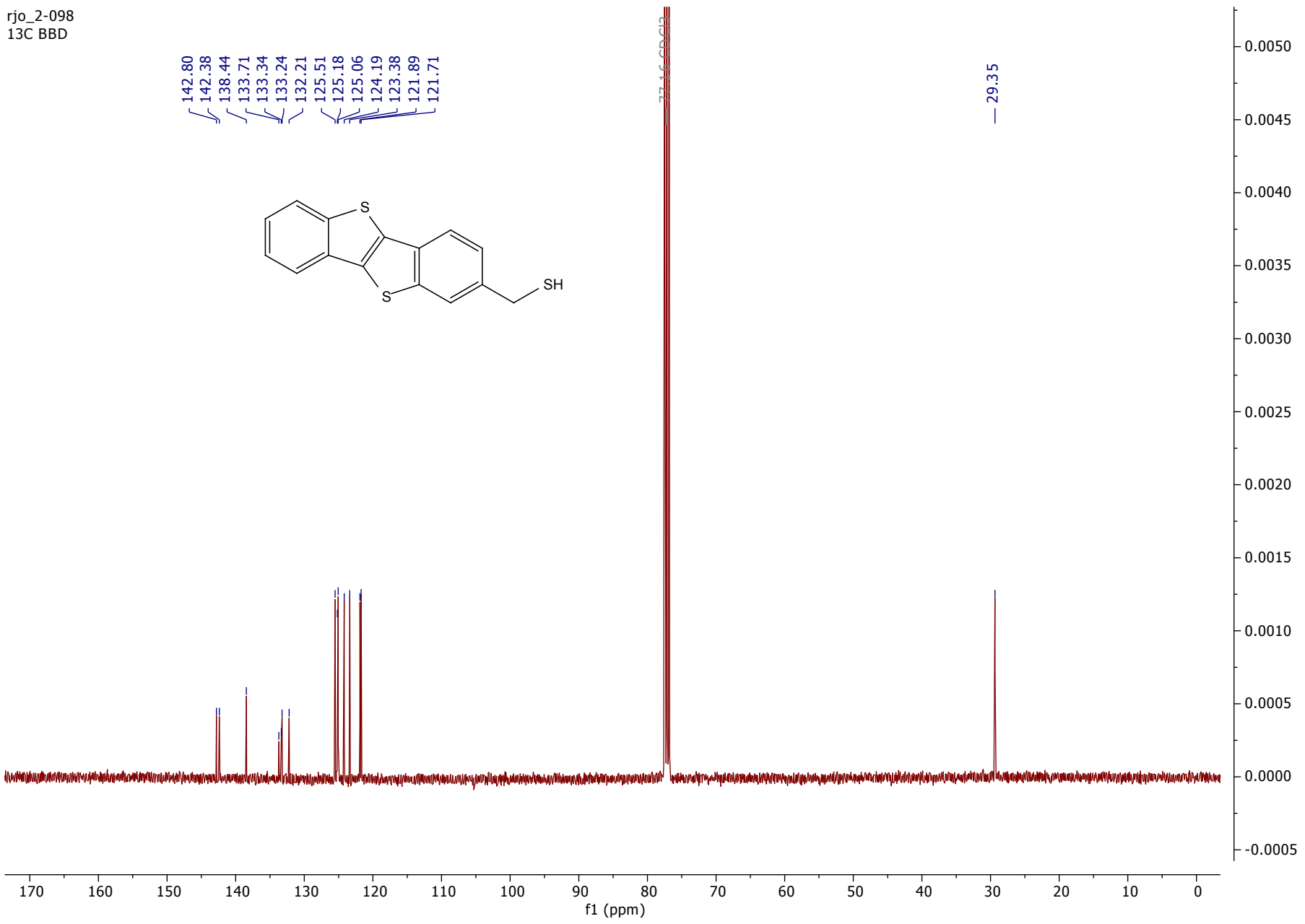

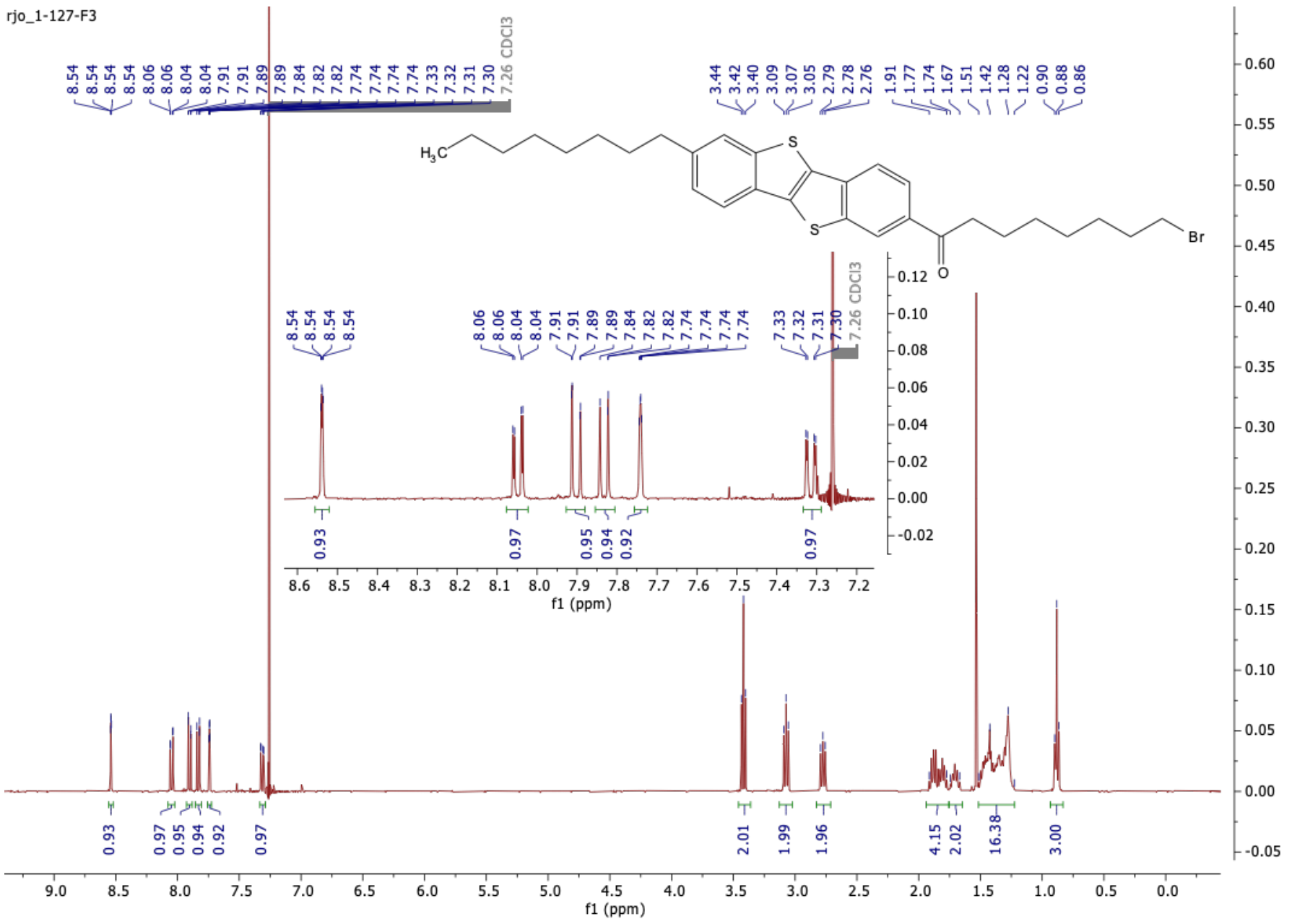

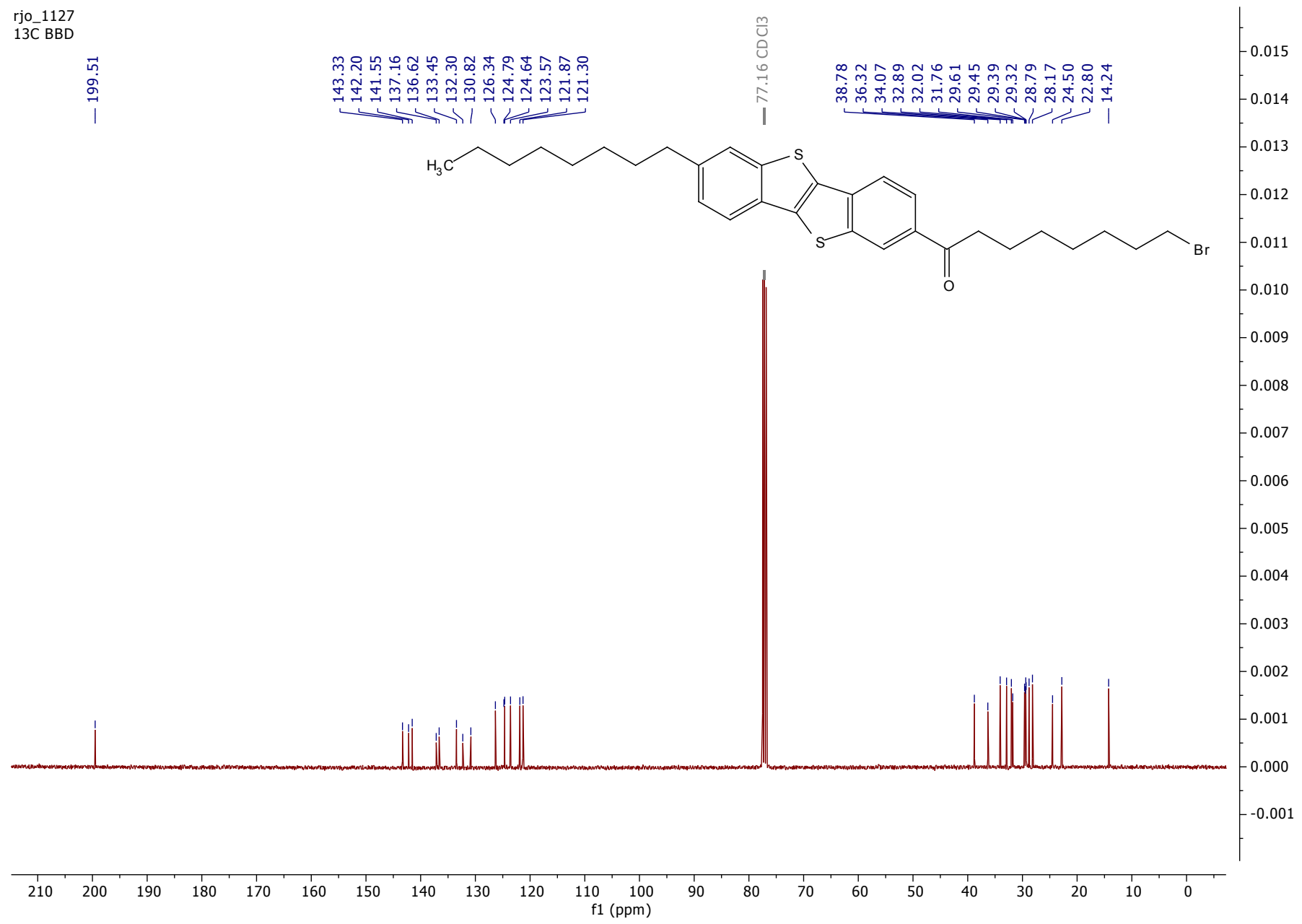

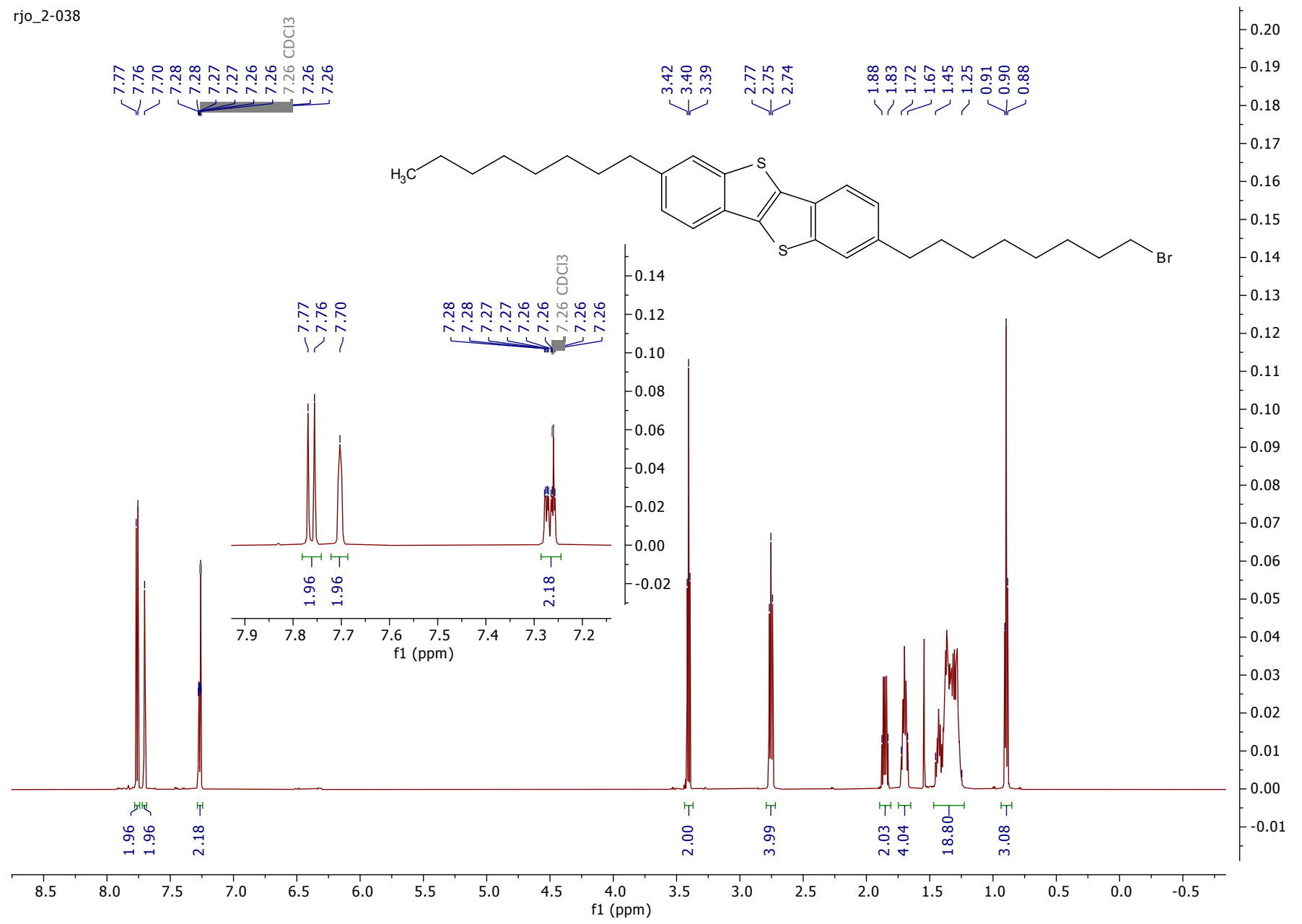

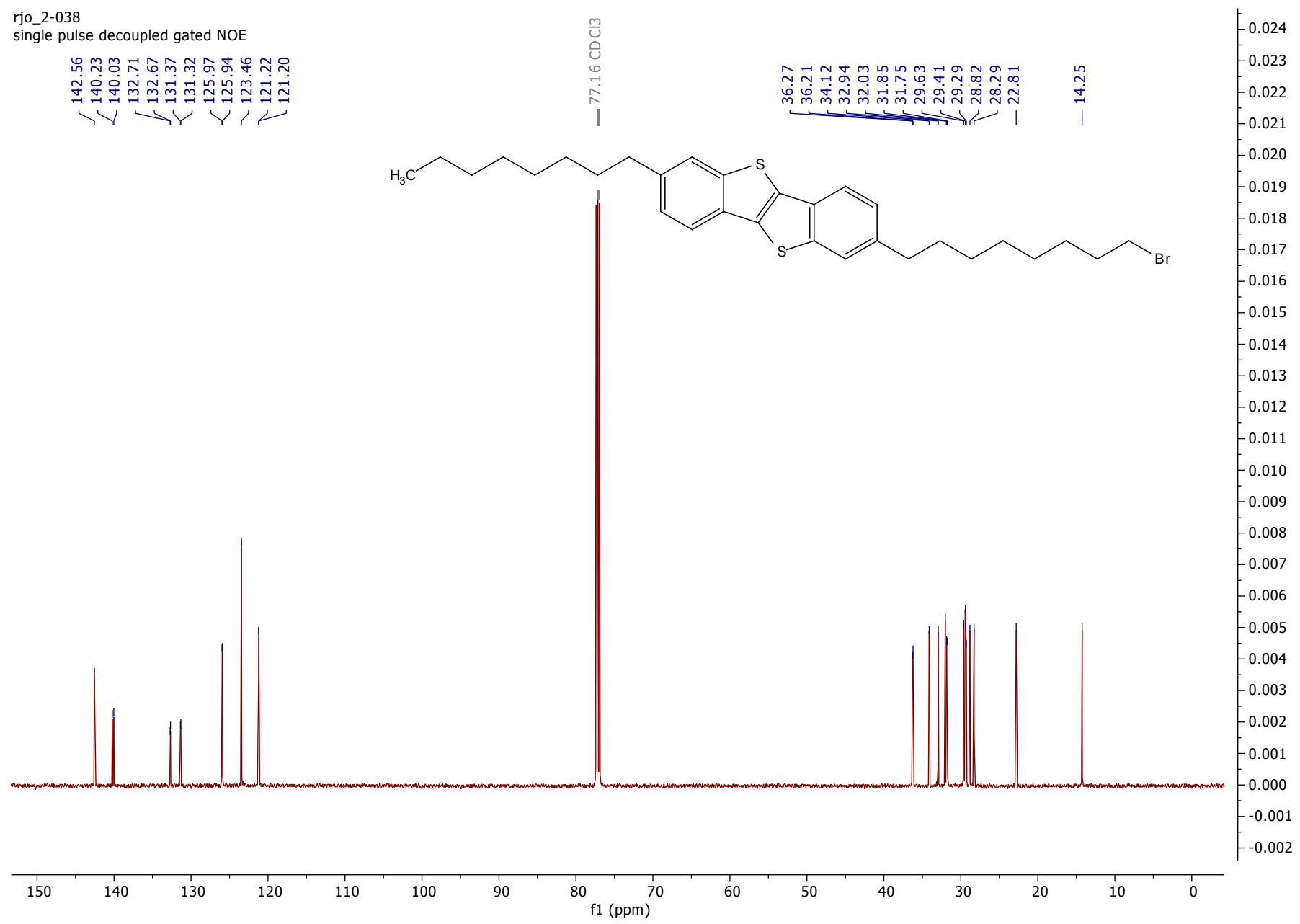

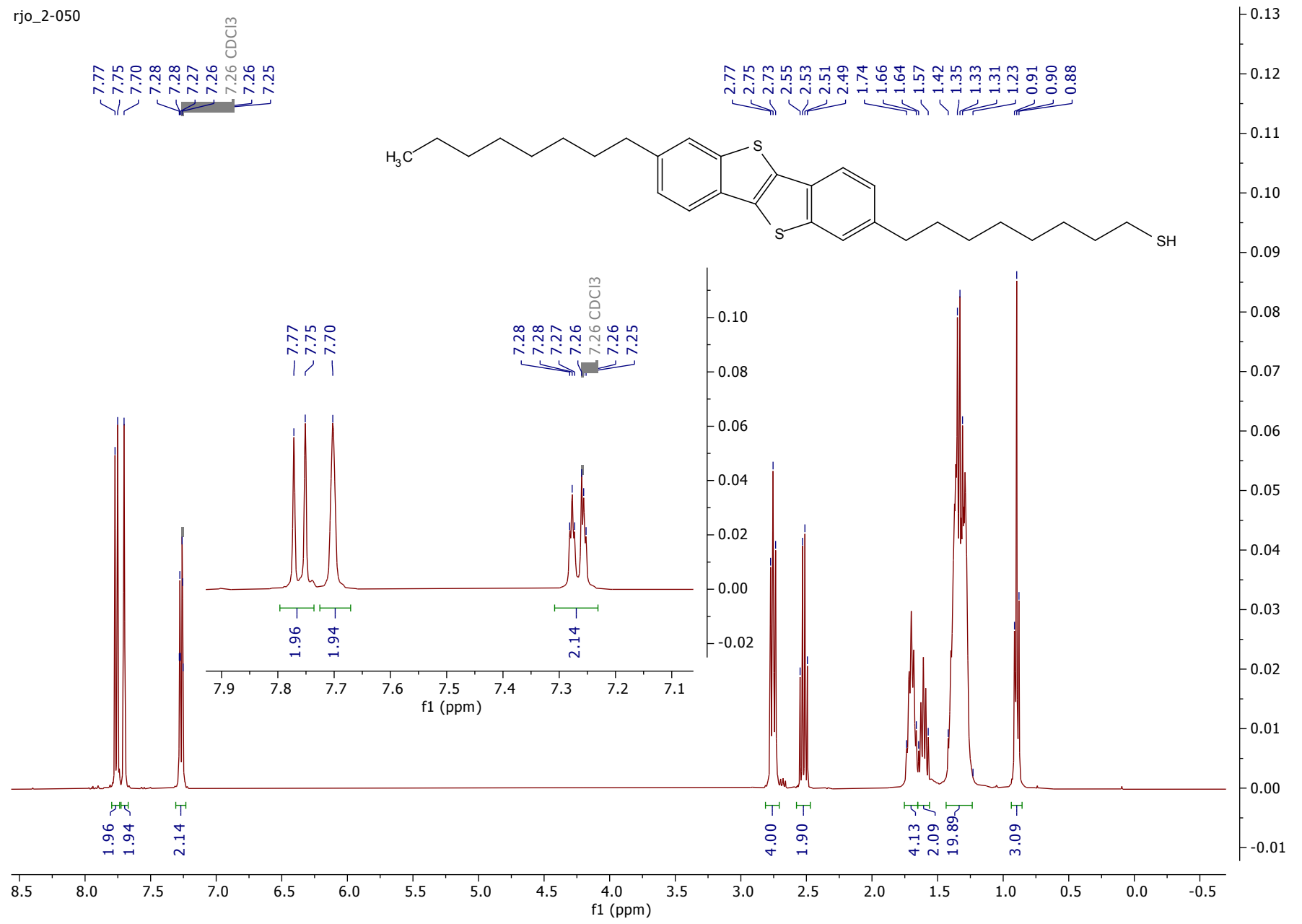

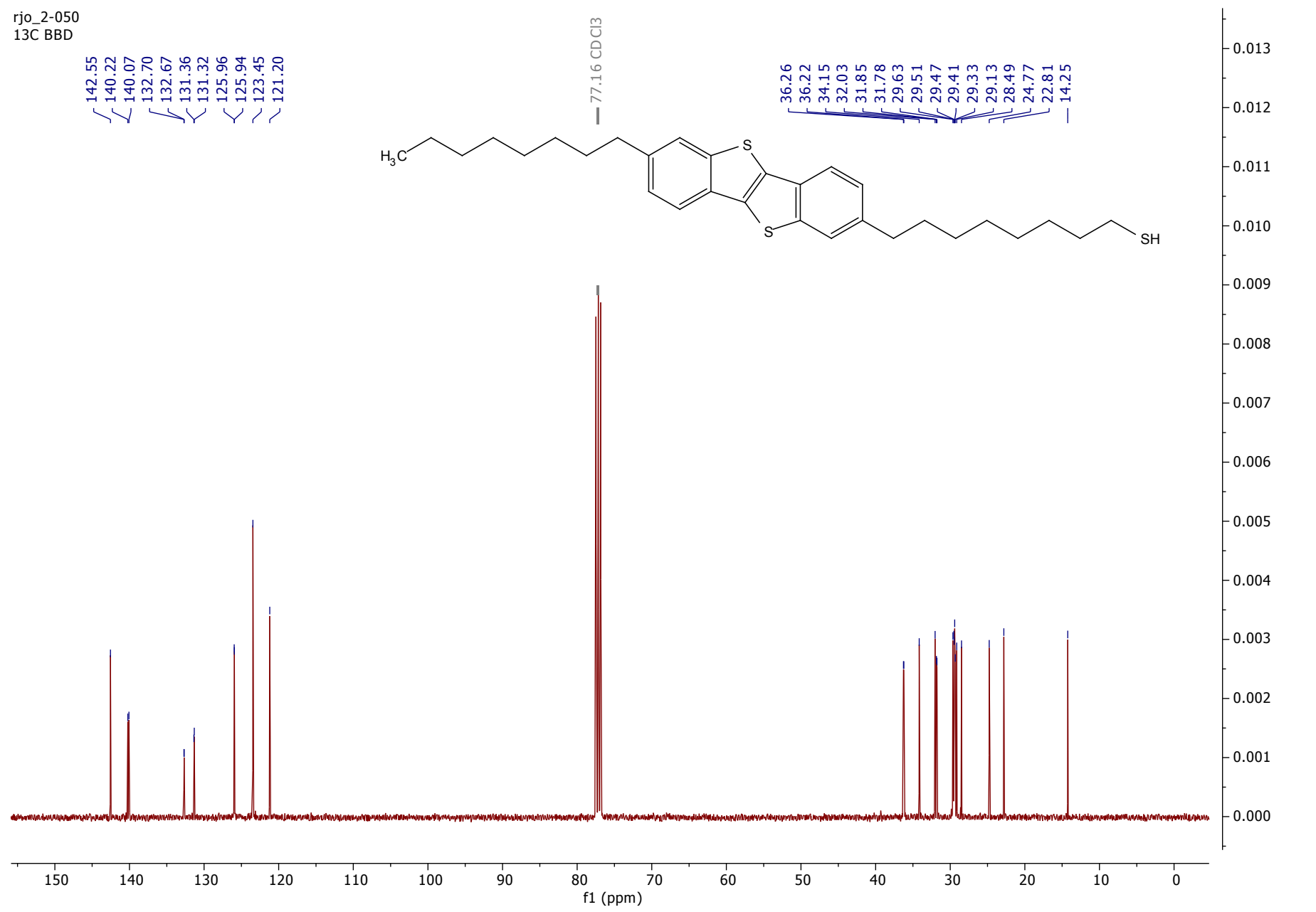

### 4. MS spectra

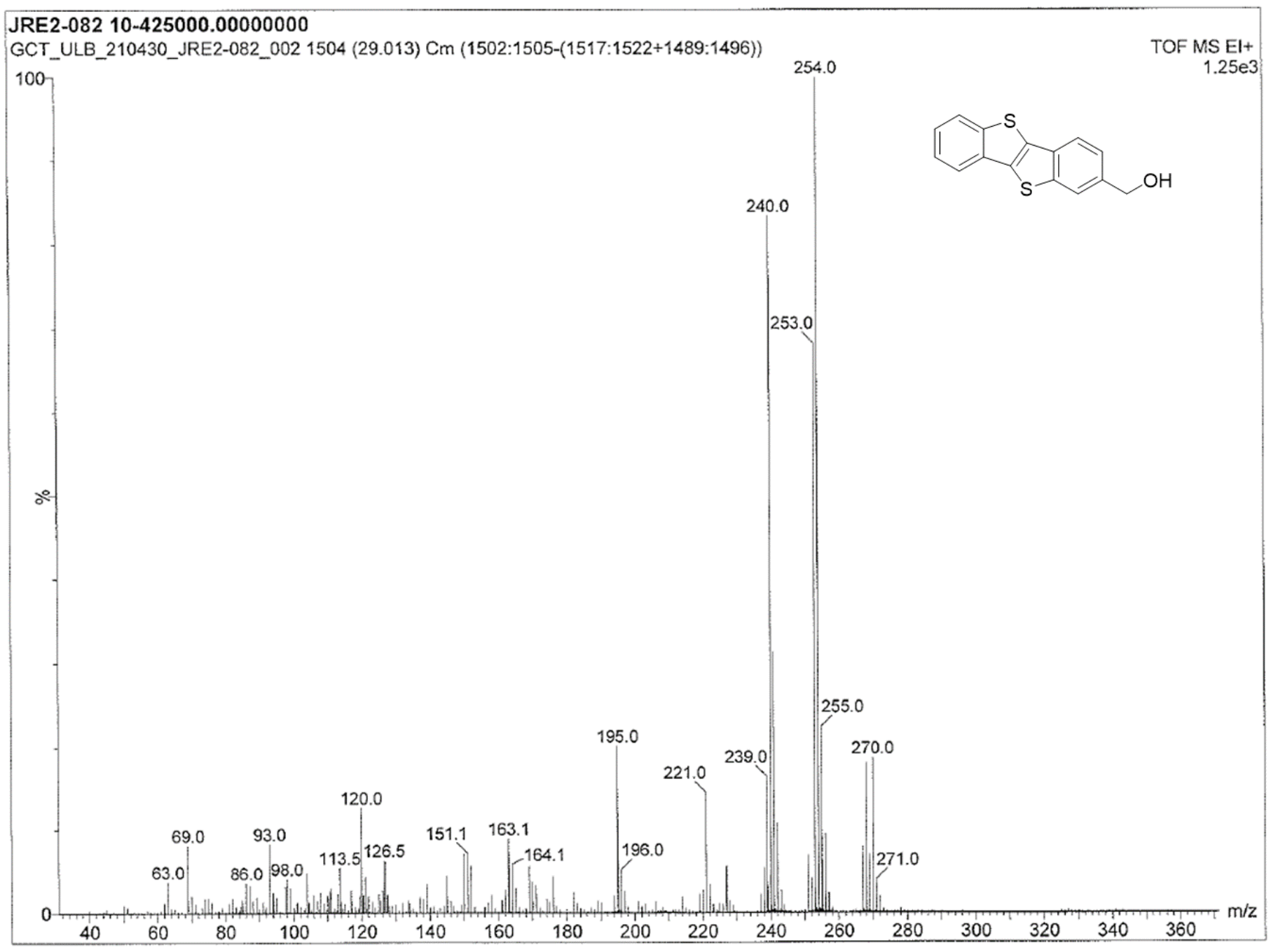

JRE2-082 10-425000.00000000
GCT_ULB_210430_JRE2-082_002 1504 (29.013) Cm (1502:1505-(1517:1522+1489:1496))
TOF MS EI+
1.25e3
254.0
240.0
253.0
255.0
195.0
270.0
239.0
221.0
120.0
151.1
163.1
69.0
93.0
196.0
126.5
113.5
164.1
271.0
63.0
86.0
98.0
S
OH
%
m/z

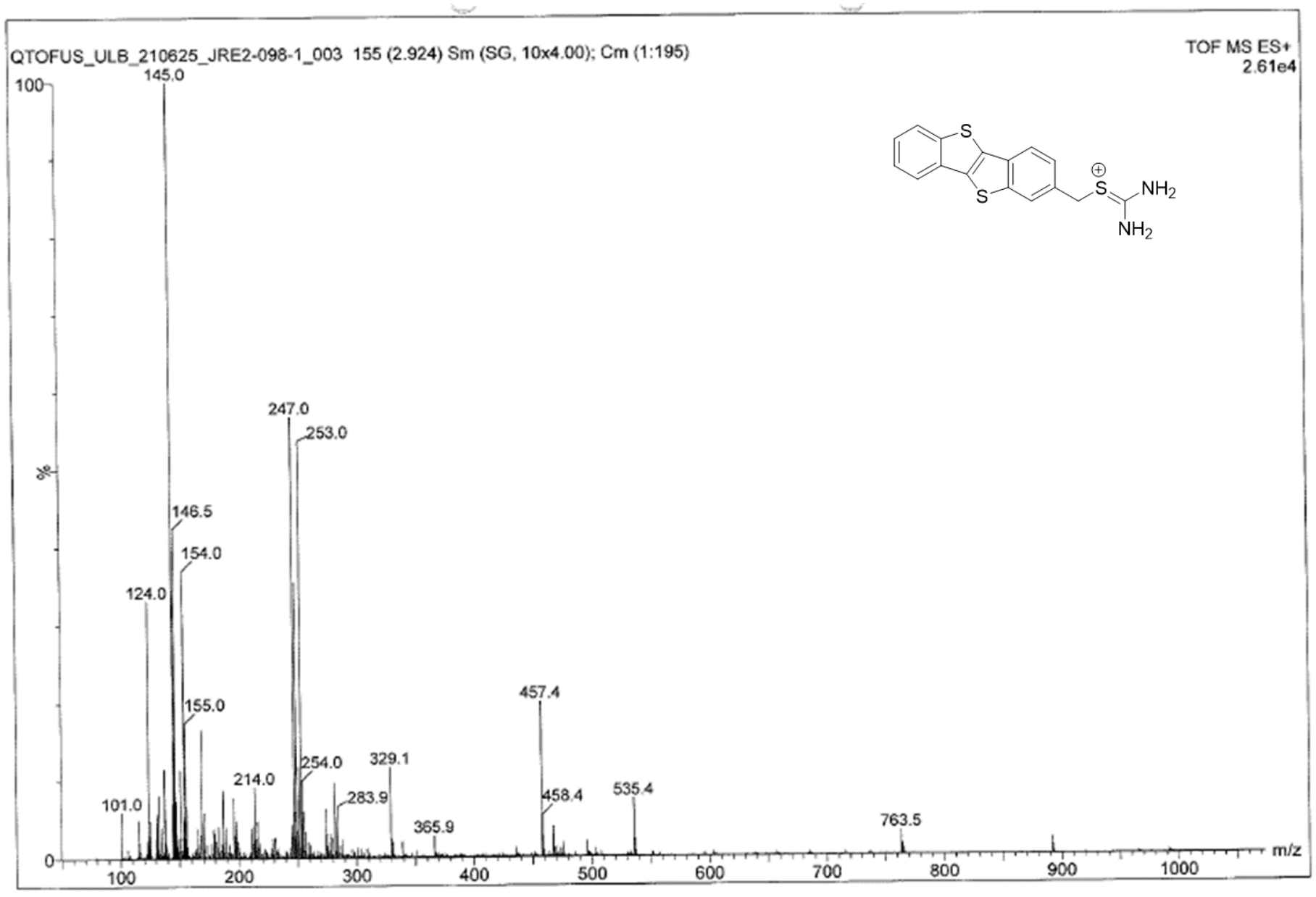

QTOFUS_ULB_210625_JRE2-098-1_003 155 (2.924) Sm (SG, 10x4.00); Cm (1:195)
TOF MS ES+
2.61e4
145.0
247.0
253.0
146.5
154.0
124.0
155.0
457.4
214.0
254.0
329.1
283.9
101.0
458.4
535.4
365.9
763.5
S
NH2
NH2
%
m/z

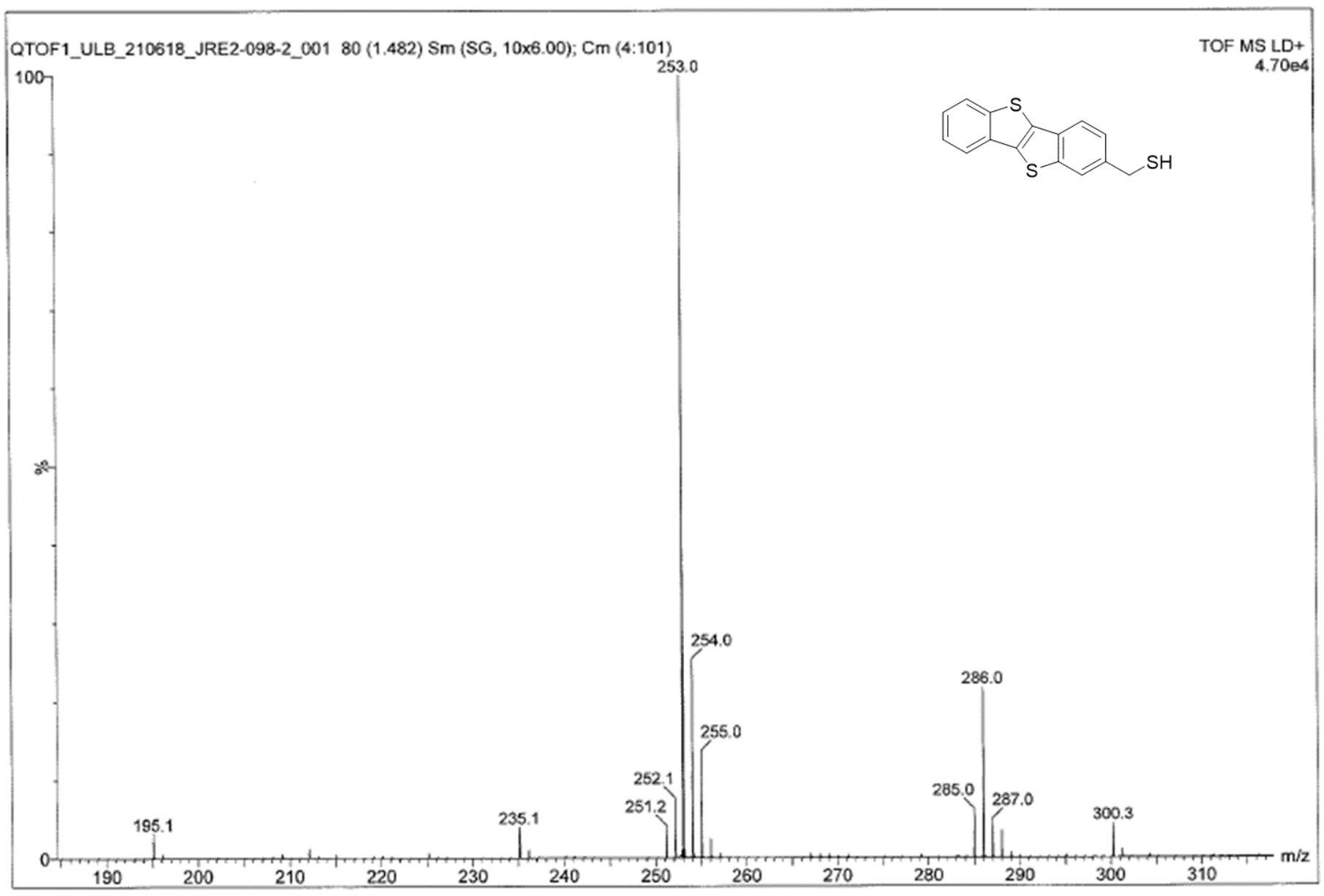

QTOF1_ULB_210618_JRE2-098-2_001 80 (1.482) Sm (SG, 10x6.00); Cm (4:101)
TOF MS LD+
4.70e4
100
%
0
253.0
254.0
255.0
252.1
251.2
286.0
285.0
287.0
300.3
235.1
195.1
190
200
210
220
230
240
250
260
270
280
290
300
310
m/z
S
SH

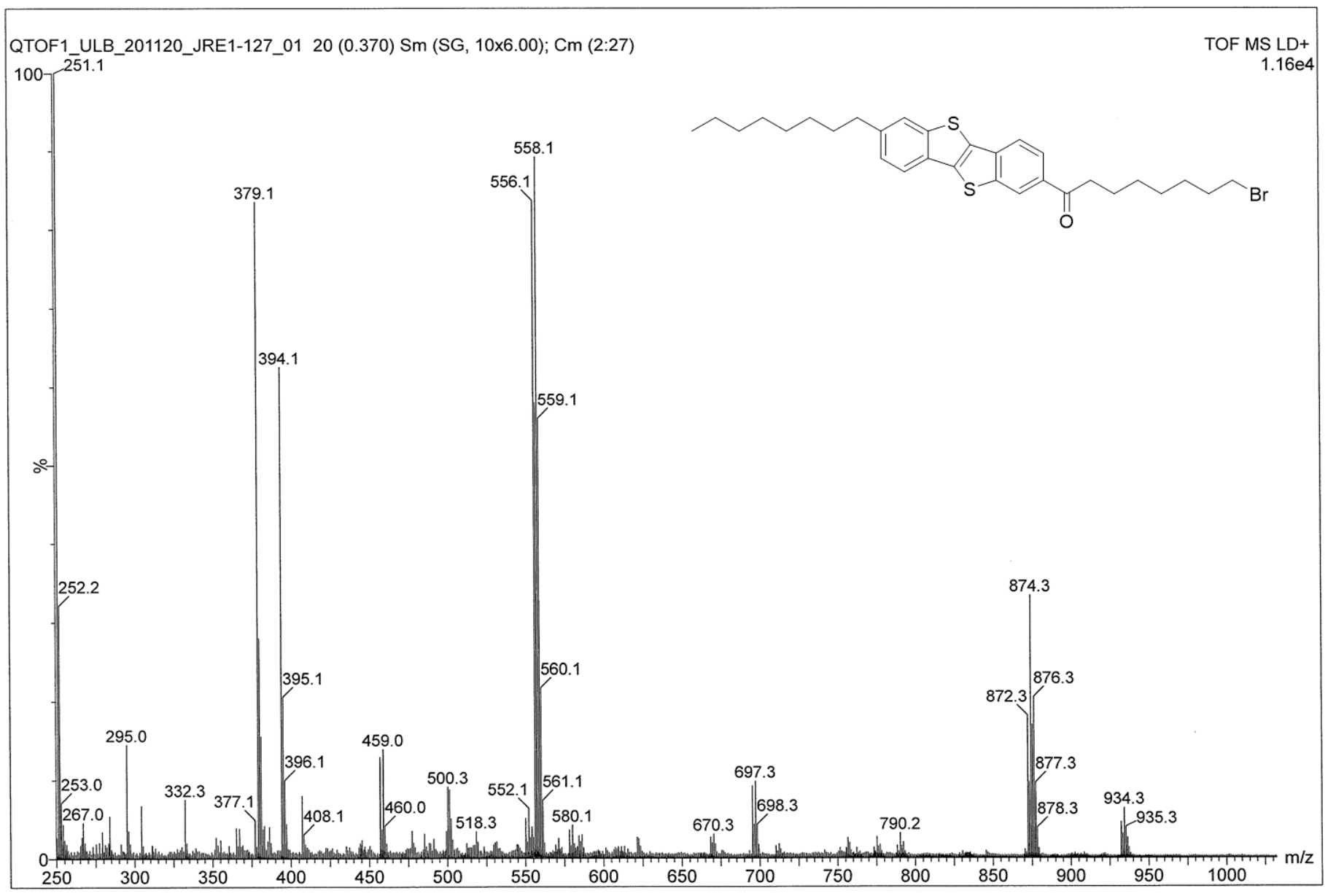

QTOF1_ULB_201120_JRE1-127_01 20 (0.370) Sm (SG, 10x6.00); Cm (2:27)
TOF MS LD+
1.16e4
100
%
0
251.1
558.1
556.1
379.1
394.1
559.1
252.2
874.3
872.3
876.3
395.1
560.1
295.0
459.0
396.1
500.3
697.3
877.3
253.0
332.3
377.1
552.1
561.1
934.3
267.0
408.1
460.0
518.3
580.1
670.3
698.3
790.2
878.3
935.3
250
300
350
400
450
500
550
600
650
700
750
800
850
900
950
1000
m/z
S
O
Br

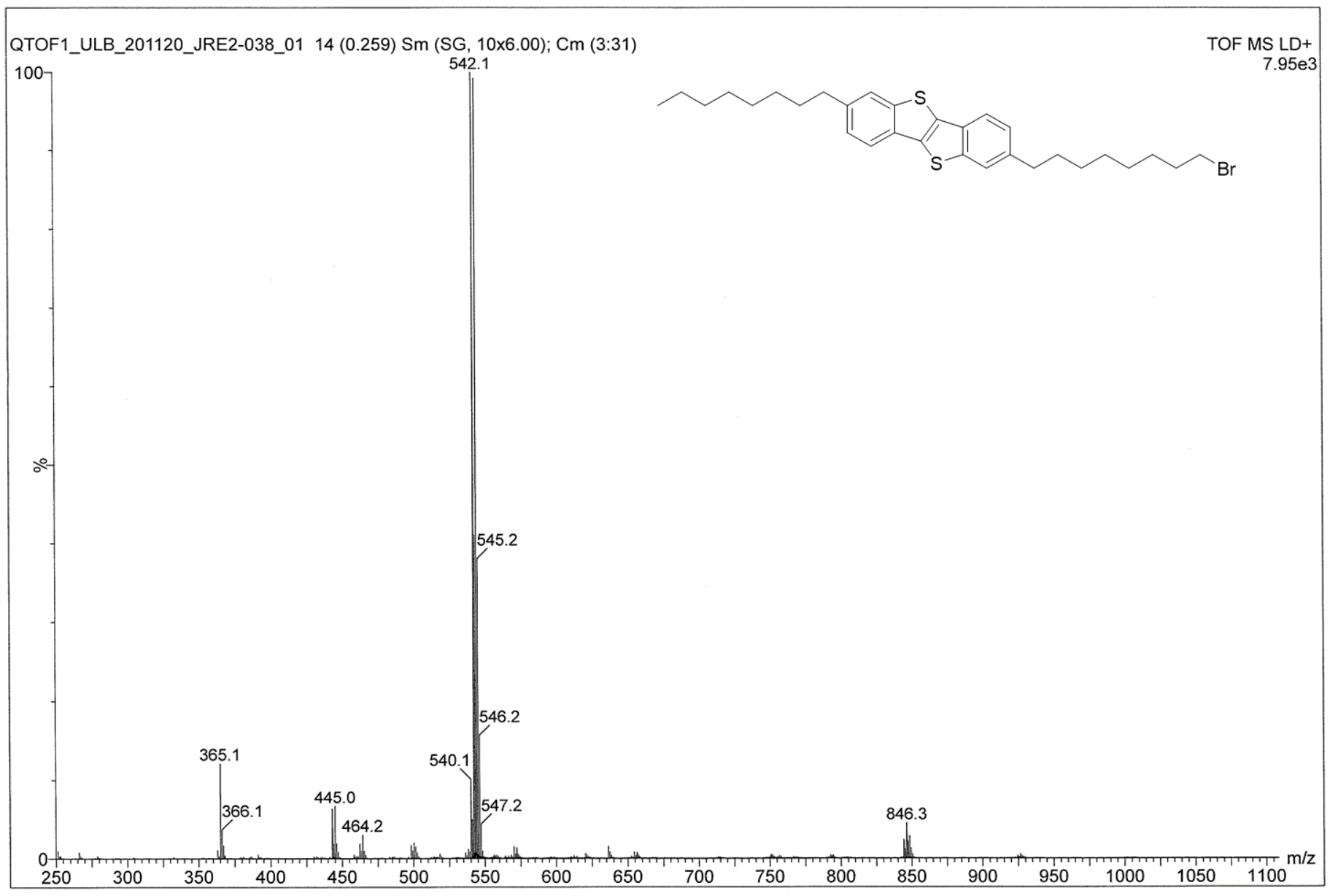

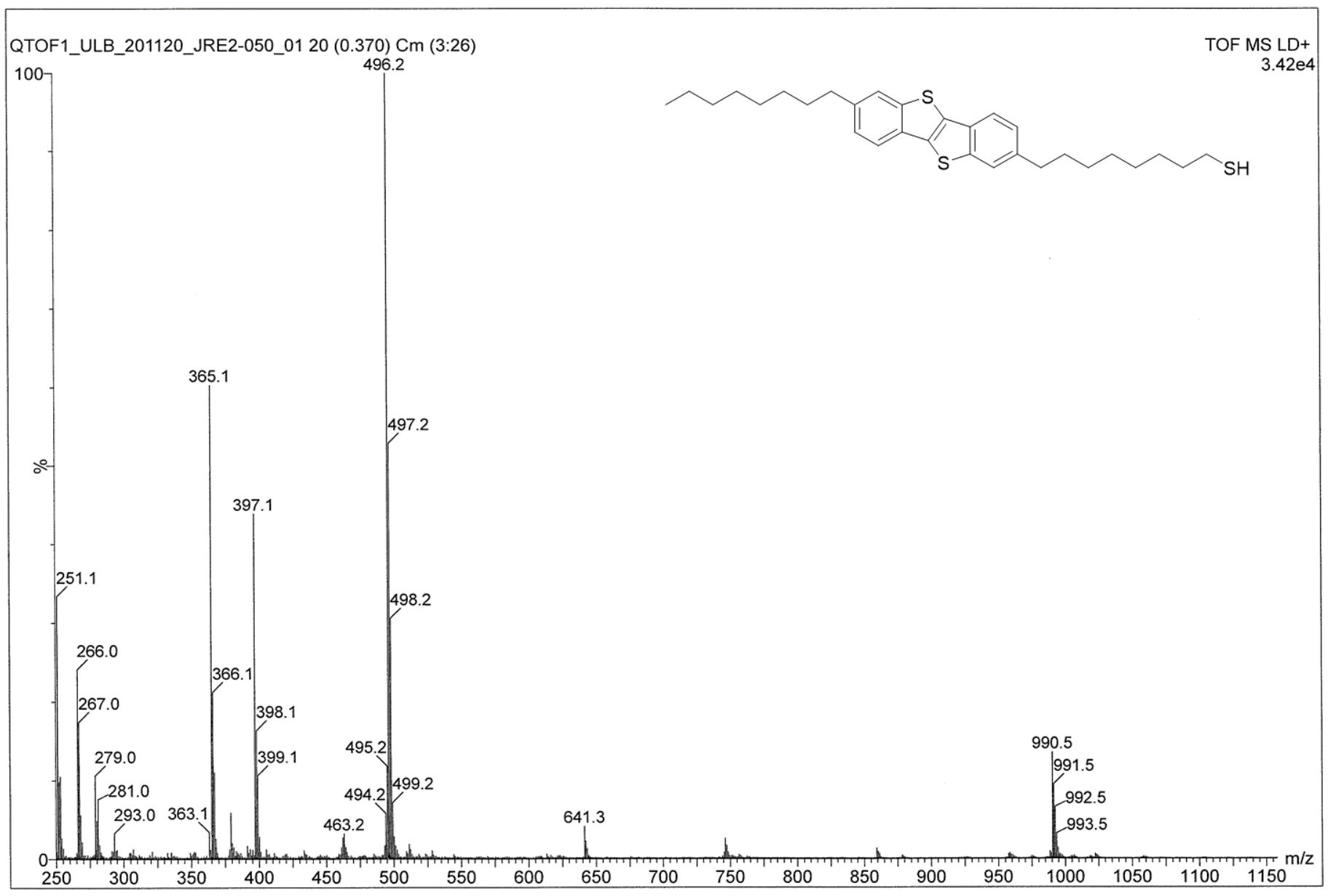

## II. Self-assembled monolayers on Au electrodes.

Ultraflat template-stripped gold surfaces ($^{TS}$Au), with rms roughness of ~0.4 nm were prepared according to methods already reported.[2-4] In brief, a 300−500 nm thick Au film was evaporated on a very flat silicon wafer covered by its native $SiO_2$ (rms roughness of ~0.4 nm), which was previously carefully cleaned by piranha solution (30 min in 7:3 $H_2SO_4/H_2O_2$ (v/v); **Caution**: Piranha solution is a strong oxidizer and reacts exothermically with organics), rinsed with deionized (DI) water, and dried under a stream of nitrogen. Clean 10x10 mm pieces of glass slide (ultrasonicated in acetone for 5 min, ultrasonicated in 2-propanol for 5 min, and UV irradiated in ozone for 10 min) were glued on the evaporated Au film (UV-polymerizable glue, NOA61 from Epotecny), then mechanically peeled off providing the $^{TS}$Au film attached on the glass side (Au film is cut with a razor blade around the glass piece).

The self-assembled monolayers (SAMs) of HS-C-BTBT and HS-$C_8$-BTBT-$C_8$ were prepared from a 1 mM solution of the molecules in a mix of THF (70%) and ethanol (30%). The $^{TS}$Au substrates were immersed for 30 hours in this solution, and then cleaned in THF with ultrasounds for one minute.

## III. Ellipsometry.

We recorded spectroscopic ellipsometry data (on ca. 1 $cm^2$ samples) in the visible range using a UVISEL (Horiba Jobin Yvon) spectroscopic ellipsometer equipped with DeltaPsi 2 data analysis software. The system acquired a spectrum ranging from 2 to 4.5 eV (corresponding to 300−750 nm) with intervals of 0.1 eV (or 15 nm). The data were taken at an angle of incidence of 70°, and the compensator was set at 45°. We fit the data by a regression analysis to a film-on-substrate model as described by their thickness and their complex refractive indexes. First, a background for the substrate before monolayer deposition was recorded. We acquired three reference spectra at three different places of the surface spaced of few mm. Secondly, after the monolayer deposition, we acquired once again three spectra at three different places of the surface and we used a 2-layer model (substrate/SAM) to fit the measured data and to determine the SAM thickness. We employed the previously measured optical properties of the substrate (background), and we fixed the refractive index of the monolayer at 1.50.[5] We note that a change from 1.50 to 1.55 would result in less than a 1 Å error for a thickness less than 30 Å. The three spectra measured on the sample were fitted separately using each of the three reference spectra, giving nine values for the SAM thickness. We calculated the mean value from this nine thickness values and the thickness incertitude corresponding to the standard deviation. Overall, we estimated the accuracy of the SAM thickness measurements at ± 2 Å.[6]

Table S1 compares the measured thicknesses to the geometry optimized length of the molecules (in gas phase, MM2 level, Chem3D). The measured thicknesses are smaller than the length (-S to terminal H). We deduce a tilt angle of ≈40° and ≈33° to the Au surface normal for the molecules in the Au-S-C-BTBT and Au-S-$C_8$-BTBT-$C_8$ SAMs, respectively. Similarly, the area per molecule is calculated from the nominal diameter of the molecules (diameter of a virtual cylinder containing the molecule, plus twice the hydrogen Van der Waals radius) and the tilt angle, which is used to estimate the number of molecules in the SAM contacted by the C-AFM and SThM tips (*vide infra*).

| | HS-C-BTBT 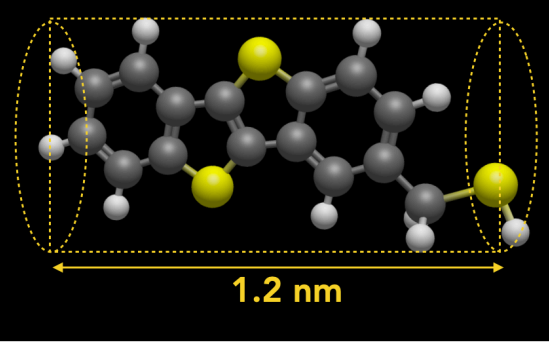  | HS-$C_8$-BTBT-$C_8$ 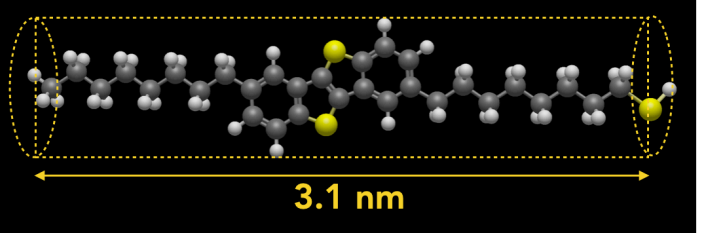  |
|---|---|---|
| Measured thickness (nm) | 0.9 ± 0.2 | 2.6 ± 0.2 |
| Theoretical length -S to H (nm) | 1.2 | 3.1 |
| Tilt angle (°) | 40 | 33 |
| Theoretical diameter (nm) | 0.58 | 0.69 |
| Projected diameter (nm) | 0.76 | 0.83 |
| Area per molecule ($nm^2$) | 0.45 | 0.53 |

***Table S1***. *Measured thickness (t) of the SAMs and comparison with the molecule length (l). The S-to-H length is taken from the geometry optimization in gas phase shown in the upper row. We also give the tilt angle Θ from the surface normal (cosΘ = t/l), the molecule diameter (d), the projected diameter on the surface (D = d/cosΘ), and the corresponding area per molecule occupied on the surface.*

## IV. Topographic AFM images.

The SAMs were examined by topographic AFM (Fig. S1) using a tip probe in silicon, model LprobeTapping20 by Vmicro. The SAM surfaces are homogeneous and flat, they are free of gross defects (neither pinhole nor aggregate, the dark spots are defects (pinholes) in the underlying Au substrate, the white spots are small dusts since the measurements were done in ambient air, both are masked for the roughness analysis). The observed rms roughness value for the surface of the $^{TS}$Au-S-C-BTBT SAM is 0.6 nm, while the value for the $^{TS}$Au-S-$C_8$-BTBT-$C_8$ SAM is 0.7 nm. Both values are close to the roughness observed for our $^{TS}$Au substrates (ca. 0.4 nm),[7] which indicates a good formation of a monolayer.

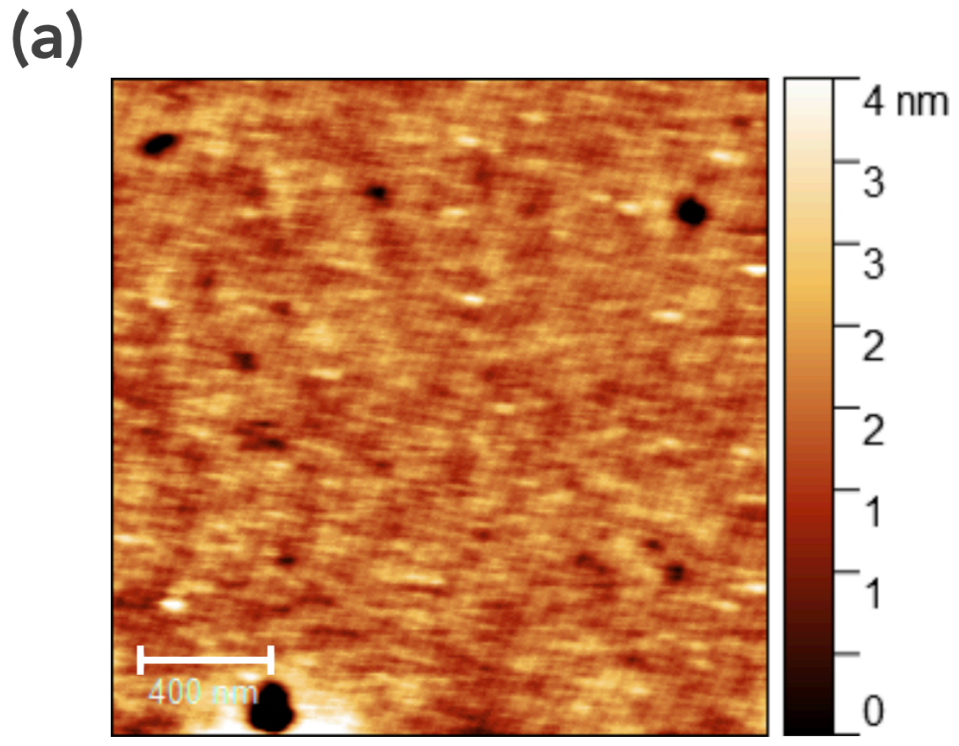

***Figure S1***. *(a) topographic AFM image of the* $^{TS}$*Au-S-C-BTBT SAM. (b) topographic AFM image of the* $^{TS}$*Au-S-*$C_8$*-BTBT-*$C_8$ *SAM. Scale bar = 400 nm.*

## V. Null-point SThM.

### *1. General procedure*

SThM[8, 9] were carried out with a Bruker ICON instrument equipped with the Anasys SThM module and in an air-conditioned laboratory (22.5°C and a relative humidity of 35-40%). We used Kelvin NanoTechnology (KNT) SThM probes with a Pd thin film resistor in the probe tip as the heating element (VITA-DM-GLA-1). The SThM tip is inserted in a Wheatstone bridge, the heat flux through the tip is controlled by the DC voltage applied on the Wheatstone bridge ($V_{WB}$, typically 0.6-1.1 V). The tip temperature, $T_{tip}$, is obtained by measuring the output voltage of the Wheatstone bridge, knowing the transfer function of the bridge, the gain of the voltage amplifier and the calibrated linear relationship between the tip resistance and the tip temperature.

The null-point SThM[10] was used at selected points on the SAMs. We define a 5x5 grid, each point spaced by 10 nm. At each point of the grid, in the z-trace mode (approach and retract) we recorded the tip temperature versus distance curve ($T_{tip}$-$z$). At the transition from a non-contact (NC, tip very near the surface) to a contact (C, tip on the surface) situation, we observe a temperature jump, $T_{NC}$ - $T_C$, which is used to determine the sample thermal conductivity according to the protocol described in Ref. 10. The temperature jump is measured from the approach trace only (to avoid any artifact due to well-known adhesion hysteresis of the retract curve) and averaged over the 25 recorded $T_{tip}$-$z$ traces. This differential method is suitable to remove the parasitic contributions (air conduction, etc…): at the contact (C) both the sample and parasitic thermal contributions govern the tip temperature, whereas, just before physical tip contact (NC), only the parasitic thermal contributions are involved. The plot of the temperature jump, $T_{NC}$ - $T_C$, versus the sample temperature at contact $T_C$ is linear and its slope is inversely proportional to the thermal conductivity. The tip-sample temperature $T_C$ increases with the supply voltage of the Wheatstone bridge $V_{DC}$ (typically from 0.6 to 1.1 V).

To determine the thermal conductivity from data like in Fig. 2 and using Eq. (1) - main text, we calibrated the null-point SThM according to the protocol in Ref. 10. The same $T_C$ vs. $T_{NC}$-$T_C$ measurements were done on two materials with well-known thermal conductivity: a glass slide (1.3 W $m^1$ $K^{-1}$) and a low-doped silicon wafer with its native oxide (150 W $m^{-1}$ $K^{-1}$). A new calibration was done for each new samples to cope with slight changes of the instrument parameters (*e.g.,* wear and tear of the tip, shift of loading force). Figure S2 shows a typical calibration curve. From a linear fit on the data (Fig. S2), we get $\alpha$ = 10.12 W $m^{-1}$ $K^{-1}$ and $\beta$ = 10.16 for the $^{TS}$Au-S-C-BTBT sample and $\alpha$ = 14.07 W $m^{-1}$ $K^{-1}$ and $\beta$ = 11.97 for the $^{TS}$Au-S-$C_8$-BTBT-$C_8$ one.

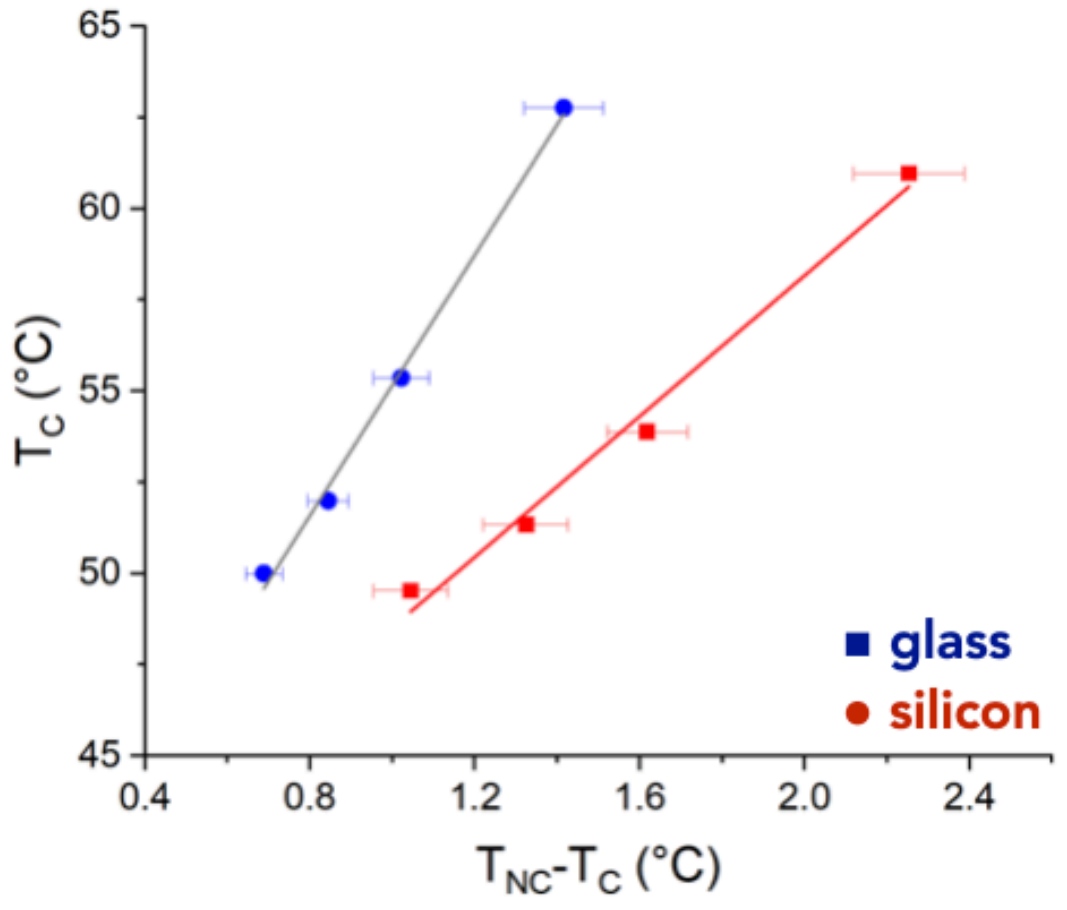

***Figure S2***. *Typical calibration curve.*

## *2. Correction with the Dryden model.*

The SAMs are deposited on a high thermal conducting Au substrate and the measured "effective" conductivity $\kappa_{SAM/Au}$ contains a contribution from the substrate. The Dryden model[11] allows calculating the constriction thermal resistance (and thus the thermal conductivity) when a small thermal spot is contacting a thin layer coating on a substrate. We used this model to calculate the effective (*i.e.* measured) conductivity of a very thin film (here the SAM) of thickness $t_{SAM}$ and thermal conductivity $\kappa_{SAM}$ deposited on a semi-infinite substrate (here the thick underlying Au electrode) with a thermal conductivity $\kappa_{Au}$ = 318 W m$^{-1}$ K$^{-1}$. In the case $t_{SAM}/r_{th}$ < 2 (here $t_{SAM}$ is 0.9 and 2.6 nm, see main text and $r_{th}$ ≈ 20 nm, *vide infra*) the model reads

$$\frac{1}{\kappa_{SAM/Au}} = \frac{1}{\kappa_{Au}} + \frac{4}{\pi\kappa_{SAM}}\left(\frac{t_{SAM}}{r_{th}}\right)\left(1-\left(\frac{\kappa_{SAM}}{\kappa_{Au}}\right)^2\right) \quad \text{(S1).}$$

Solving this equation for all the measured $\kappa_{SAM/Au}$ allows determining the SAM thermal conductivity $\kappa_{SAM}$.

### 3. Thermal contact area.

The thermal contact radius (at the tip/SAM interface) is calculated following the approach reported in Ref. 12 taken into account the mechanical tip radius $r_{tip}$ and the size of the water meniscus at the tip/surface interface. The thermal radius of the thermal contact is estimated by[13]

$$r_{th} = 2.08\sqrt{\frac{-r_{tip}\cos\theta}{\ln\varphi}} \quad \text{(S2)}$$

with $r_{tip}$ = 100 nm (data from Bruker), the relative humidity $\varphi$ = 0.35-0.4 (in an air-conditioned laboratory, values checked during the measurements) and the contact angle of the concave meniscus between the tip and the surface $\theta$ ≈ 30° as measured for π-conjugated molecular crystals in Ref. 14. We get $r_{th}$ ≈ 20 nm. The water meniscus contact angle depends on the surface energy of the sample, and thus should, in principle, not be the same for the $^{TS}$Au-S-C-BTBT SAM and $^{TS}$Au-S-$C_8$-BTBT-$C_8$ SAM, the latter being more hydrophilic due to the alkyl chains on the upper part of the SAM. However, we cannot perform water contact angle measurements inside the nanometer size tip/SAM interface, and we consider the same literature value of 30° in both cases. Considering an area per molecule of ~0.45 nm$^2$, we estimate that ~ 2500 molecules are contacted during the SThM measurements.

## VI. Conductive-AFM.

### 1. General procedure

We measured the electron transport properties at the nanoscale by C-AFM (ICON, Bruker) at room temperature (in an air-conditioned laboratory: 22.5°C and a relative humidity of 35-40%) using a tip probe in platinum/iridium (PtIr), model SCM-PIC-V2 from Bruker. We used a "blind" mode to measure the current-voltage (*I-V*) curves and the current histograms: a square grid of 10×10 was defined with a pitch of 50 to 100 nm. At each point, one *I-V* curve is acquired leading to the measurements of 100 traces per grid. This process was repeated several times at different places (randomly chosen) on the sample, and up to several thousands of *I-V* traces were used to construct the current-voltage histograms (shown in Fig. 4, main text).

The tip load force was set at ≈ 10 nN for all the *I-V* measurements, a lower value leading to too many contact instabilities during the *I-V* measurements. Albeit larger than the usual load force (2-5 nN) used for CAFM on SAMs, this value is below the limit of about 60-70 nN at which the SAMs start to suffer from severe degradations. For example, a detailed study (Ref. 15) showed a limited strain-induced deformation of the monolayer (≲ 0.3 nm) at this used load force. The same conclusion was confirmed by our own study comparing mechanical and electrical properties of alkylthiol SAMs on flat Au surfaces and tiny Au nanodots.[16]

## *2. C-AFM contact area.*

Considering: (i) the area per molecule on the surface (as estimated for the thickness measurement and calculated geometry optimization - see Table S1), and (ii) the estimated C-AFM tip contact surface (see below), we estimated the C-AFM tip contact area and the number, N, of molecules contacted by the tip as follows. As usually reported in literature[15, 17-19] the contact radius, *a*, between the C-AFM tip and the SAM surface, and the SAM elastic deformation, *δ*, are estimated from a Hertzian model:[20]

$$a^2 = \left(\frac{3RF}{4E^*}\right)^{2/3} \qquad \text{(S3)}$$

$$\delta = \left(\frac{9}{16R}\right)^{1/3}\left(\frac{F}{E^*}\right)^{2/3} \qquad \text{(S4)}$$

with *F* the tip load force (10 nN), *R* the tip radius (20 nm) and *E** the reduced effective Young modulus defined as:

$$E^* = \left(\frac{1}{E^*_{SAM}} + \frac{1}{E^*_{tip}}\right)^{-1} = \left(\frac{1-\nu^2_{SAM}}{E_{SAM}} + \frac{1-\nu^2_{tip}}{E_{tip}}\right)^{-1} \qquad \text{(S5)}$$

In this equation, $E_{SAM/tip}$ and $\nu_{SAM/tip}$ are the Young modulus and the Poisson ratio of the SAM and C-AFM tip, respectively. For the Pt/Ir (90%/10%) tip, we have $E_{tip}$ = 204 GPa and $\nu_{tip}$ = 0.37 using a rule of mixture with the known material data.[21] These parameters for the BTBT derivative SAMs are not known and, in general, they are not easily determined in such a monolayer material. Thus, we consider the value of an effective Young modulus of the SAM $E^*_{SAM}$ = 38 GPa as determined for the "model system" alkylthiol SAMs from a combined mechanic and electron transport study.[15] With these parameters, we estimated $a$ = 1.7 nm (contact area = 8.6 nm$^2$) and $\delta$ = 0.14 nm. With a molecular packing density of 0.45 nm$^2$/molecule for the HS-C-BTBT molecules (as estimated from the tilt angle and theoretical configuration optimization, see table S1), we infer that about 15 molecules (see www.packomania.com) are measured in the SAM/PtIr junction.

### 3. *Data analysis.*

Before to construct the current histograms shown in Figs. 4-a and 4-b (main text) and analyze the *I-V* curves with the one energy-level model, the raw set of *I-V* data is scanned and some *I-V* curves were discarded from the analysis:

- At high current, the *I-V* traces that reached the saturating current during the voltage scan (the compliance level of the trans-impedance amplifier, typically 5x10$^{-8}$ A here (depending on the gain of the amplifier) and/or *I-V* traces displaying large and abrupt steps during the scan (contact instabilities).

- At low currents, the *I-V* traces that reached the sensitivity limit (almost flat *I-V* traces and noisy *I-V*s) and displayed random staircase behavior (due to the sensitivity limit - typically 2-3 pA depending on the used gain of the trans-impedance amplifier and the resolution of the ADC (analog-digital converter).

Figure S3 shows the 2D histograms before this data analysis (made with all the acquired *I-V* traces).

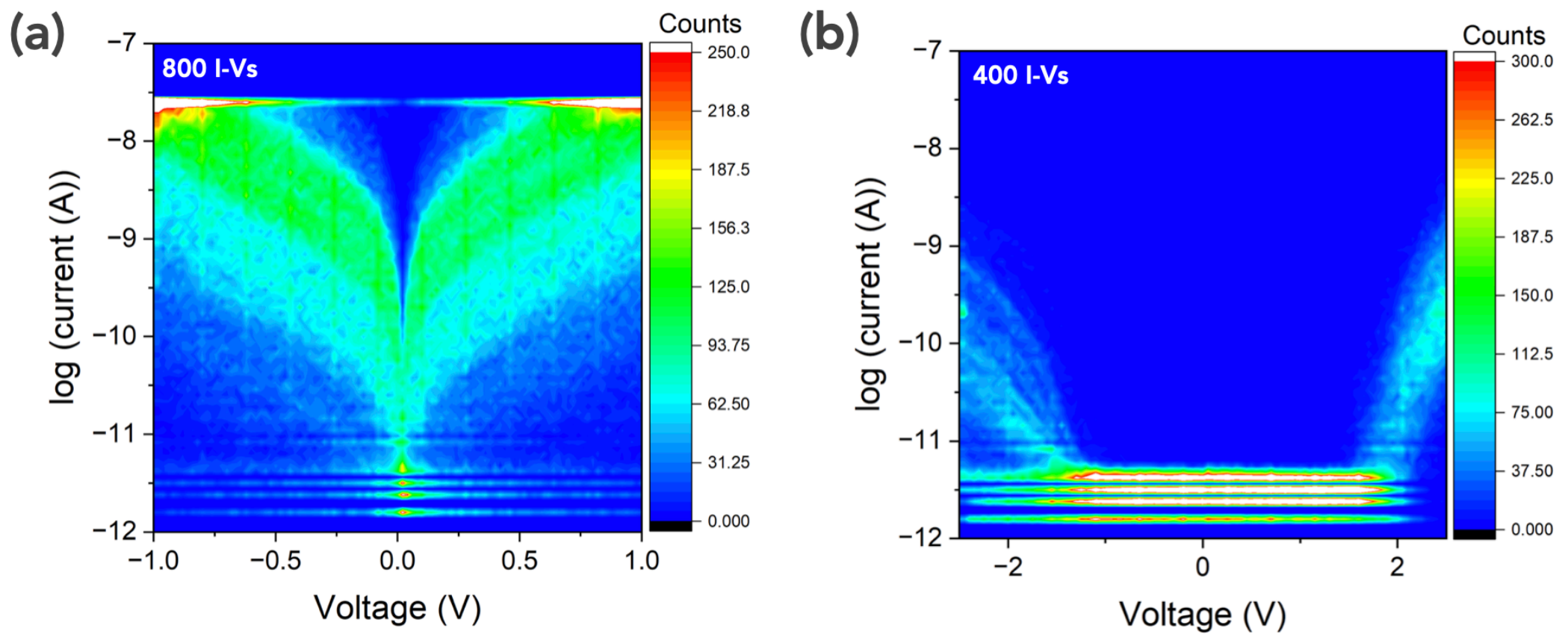

***Figure S3.*** *2D histograms of the complete I-V dataset: (a) $^{TS}$Au-S-C-BTBT/PtIr C-AFM tip molecular junctions, 800 I-V traces were acquired; (b) $^{TS}$Au-S-$C_8$-BTBT-$C_8$/PtIr C-AFM tip molecular junctions, 400 I-Vs traces were acquired.*

### 4. Statistical analysis of the I-V dataset.

Figure S4 shows the statistical distribution of the currents (absolute value) of the $^{TS}$Au-S-C-BTBT/PtIr C-AFM tip molecular junction measured at -0.5 and 0.5V (from the dataset in Fig. 4-a). The currents are broadly distributed and they can be fitted with two log-normal distributions for the HC (high current) and LC (low current) parts. The log-mean (log-μ) and log-standard deviation (log-σ) are given in the figure caption. The HC and LC contributions correspond to MJs with the lowest energy level $\varepsilon_{01}$ and highest $\varepsilon_{02}$ (Fig. 4-c, main text), respectively.

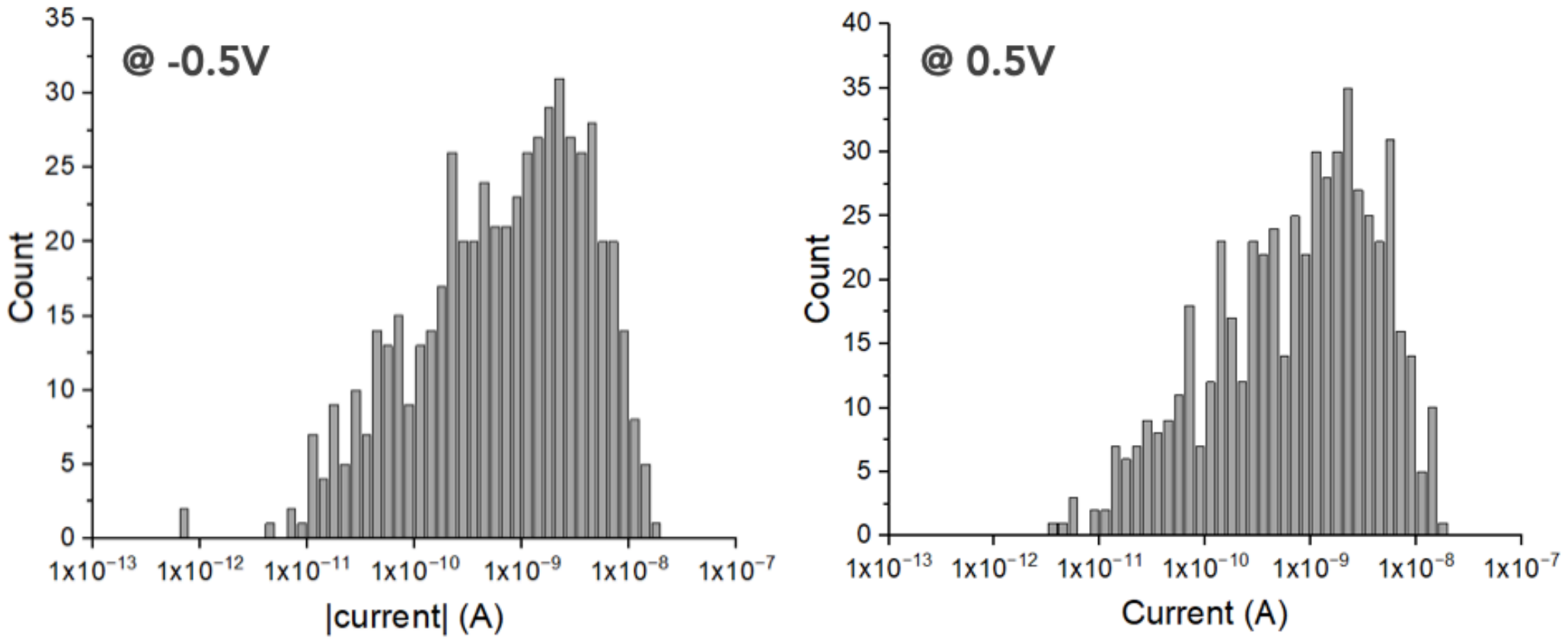

***Figure S4**. Statistical distribution of the currents (at -0.5 and 0.5 V) for the $^{TS}$Au-S-C-BTBT/PtIr C-AFM tip molecular junctions.*

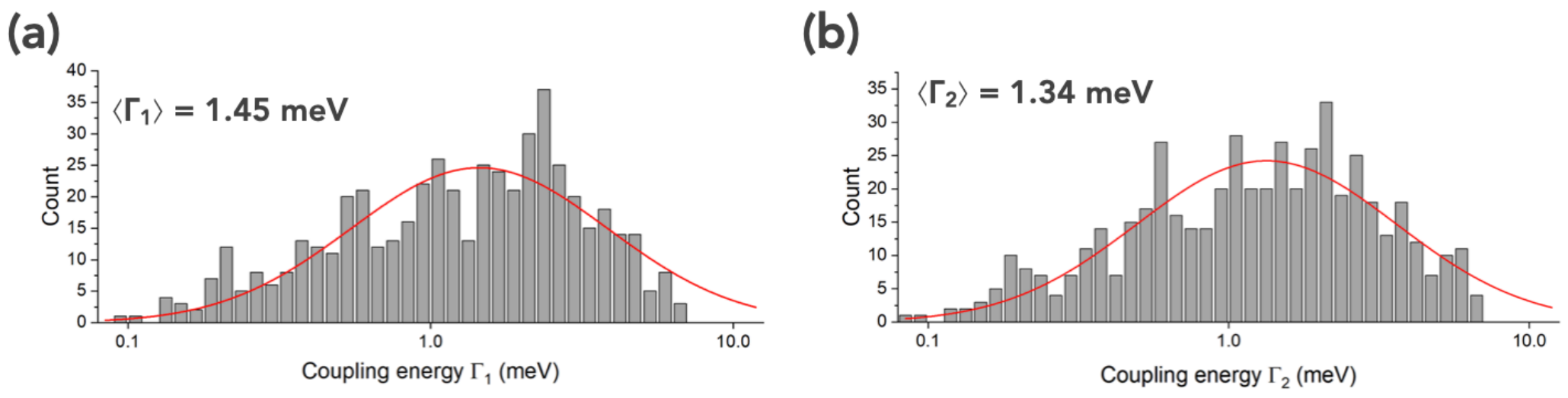

***Figure S5**. Statistical distribution of the electrode coupling energies $\Gamma_1$ and $\Gamma_2$ for the I-V dataset of $^{TS}$Au-S-C-BTBT/PtIr C-AFM tip MJs shown Fig. 4-a. The red lines are fits by a log-normal distribution*

*with log-mean, log-μ=0.161 (or mean $\langle \Gamma_1 \rangle$=1.45 meV), log standard deviation, log-σ=0.425, for $\Gamma_1$ and log-μ=0.127 (or mean $\langle \Gamma_2 \rangle$=1.34 meV), log-σ=0.433 for $\Gamma_2$.*

## 5. Fits of the I-V curves with the analytical SEL model.

All the I-V traces of the dataset of the $^{TS}$Au-S-C-BTBT/PtIr C-AFM tip molecular junctions (Fig. 4-a) were fitted with the single-energy level (SEL) model given by the following analytical expression:[22, 23]

$$I(V) = N\frac{8e}{h}\frac{\Gamma_1\Gamma_2}{\Gamma_1+\Gamma_2}\left[arctan\left(\frac{\varepsilon_0+\frac{\Gamma_1}{\Gamma_1+\Gamma_2}eV}{\Gamma_1+\Gamma_2}\right)-arctan\left(\frac{\varepsilon_0-\frac{\Gamma_2}{\Gamma_1+\Gamma_2}eV}{\Gamma_1+\Gamma_2}\right)\right] \quad \text{(S6)}$$

with $\varepsilon_0$ the energy of the molecular orbital (MO), here HOMO, involved in the transport (with respect to the Fermi energy of the electrodes), $\Gamma_1$ and $\Gamma_2$ the electronic coupling energy between the MO and the electron clouds in the two electrodes, *e* the elementary electron charge, *h* the Planck constant and *N* the number of molecules contributing to the ET in the molecular junction (assuming independent molecules conducting in parallel, *i.e.* no intermolecular interaction[24-26]). We used *N* = 15, *vide supra*.

This model is valid at 0 K, since the Fermi-Dirac electron distribution of the electrodes is not taken into account. However, it was shown that it can be reasonably used to fit data measured at room temperature for voltages below the transition between the off-resonant and resonant transport conditions at which the broadening of the Fermi function modify the I-V shape leading to sharpened increase of the current.[27-29] We defined this bias voltage window of confidence by TVS (transition voltage spectroscopy) that give us an estimate of this transition regime. Figure S5-a shows the TVS curves obtained from the mean $\bar{I}$-$V$ of the $^{TS}$Au-S-$C_8$-BTBT-$C_8$/PtIr C-AFM tip molecular junction. The maxima (red arrows) indicate transition voltages at $V_{T-}$=-0.72 V and $V_{T+}$≈0.83 V. Therefore, we fixed the bias window of confidence between - 0.8 V and 0.8 V to fit all the *I-Vs* of the dataset. Figure S5-b shows a typical fit of the SEL model (Eq. S6) on the mean $\bar{I}$-$V$ of the $^{TS}$Au-S-C-BTBT/PtIr C-AFM tip molecular junctions. We note that, with these conditions, the two methods give almost the same value for the energy level $\varepsilon_0$ (0.62 eV fit with SEL, 0.67 eV by TVS, *vide infra* equation S7). We also verified *a posteriori* that the condition of applicability of the 0K SEL model to room temperature experimental data is satisfied by using a numerical analysis reported in Ref. 30. In our case, with $\varepsilon_0$ ≈ 0.5-07 eV, $\Gamma_1$ and $\Gamma_2$ around 1 meV, this condition is |V|< 0.64 - 1 V). The fits of the SEL model were done with the routine included in ORIGIN software

(version 2023, OriginLab Corporation, Northampton, MA, USA), using the method of least squares and the Levenberg Marquardt iteration algorithm.

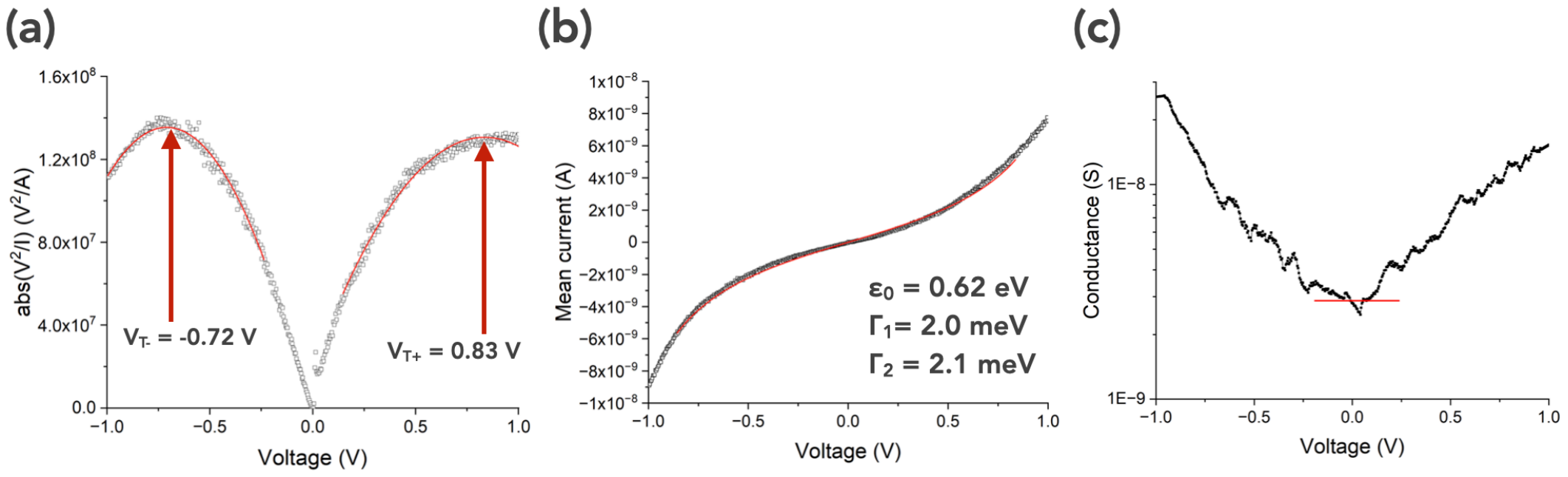

***Figure S5***. ***(a)*** *Data of the mean Ī-V of the* $^{TS}$*Au-S-C-BTBT/PtIr C-AFM tip molecular junction plotted as* $|V^2/I|$ *vs. V. The red lines are fits by a 2nd order polynomial function.* ***(b)*** *Fit of the SEL model (Eq. S6) on the mean Ī-V of the* $^{TS}$*Au-S-C-BTBT/PtIr C-AFM tip molecular junction, fit limited between -0.8 and 0.8 V, the fit parameters are given in the panel.* ***(c)*** *Conductance (1$^{st}$ derivative) of the data in panel b, the horizontal red line indicates the value of the zero bias conductance: 2.7x10$^{-9}$ S.*

We also used the transition voltage spectroscopy (TVS)[31-35] to analyze the I-V curves. Plotting $|V^2/I|$ *vs. V* (Fig. S4-a),[36] we determine the transition voltages $V_{T+}$ and $V_{T-}$ for both voltage polarities at which the bell-shaped curve is maximum. This threshold voltage indicates the transition between off resonant (below $V_T$) and resonant (above $V_T$) transport regime in the molecular junctions and can therefore be used to estimate the location of the energy level. In Fig. S5-a, the thresholds $V_{T+}$ and $V_{T-}$ are indicated by the vertical arrows (with values) and determined from the max of a 2nd order polynomial function fitted around the max of the bell-shaped curves (to cope with noisy curves). The value of $\varepsilon_{0\text{-}TVS}$ is estimated by:[35]

$$\left|\varepsilon_{0-TVS}\right| = 2\frac{e\left|V_{T+}V_{T-}\right|}{\sqrt{V_{T+}^2 + 10\left|V_{T+}V_{T-}\right|/3 + V_{T-}^2}} \qquad \text{(S7)}$$

The zero-bias conductance of the $^{TS}$Au-S-C-BTBT/PtIr C-AFM tip molecular junction was calculated from the first derivative of the I-V curves. A typical example for the mean Ī-V is shown in Fig. S5-c. The numerical derivative was smoothed with the Savitzky-Golay[37] digital filter using a second order polynomial function and 20 points of window. We estimated the conductance per molecule simply by dividing the SAM conductance by the number of molecules under the C-AFM tip junction (15, *vide supra*) assuming independent molecules conducting in parallel, *i.e.* no intermolecular interaction.[24-26]

## VII. Thermal equivalent circuit.

The figure S6 shows a simple thermal equivalent circuit of the two MJs. $R_{c1}$ and $R_{c2}$ are the contact thermal resistances at the tip and bottom Au electrodes, respectively. They are not known and we assume that they have the same values in both the cases. $R_{BTBT}$ stands for the thermal resistance of the BTBT moiety (assume to be the same in the two MJs) and $R_{C8}$ is the thermal resistance of the alkyl chains. Thus, the difference between the measured thermal conductances give an estimation of $2R_{C8}$.

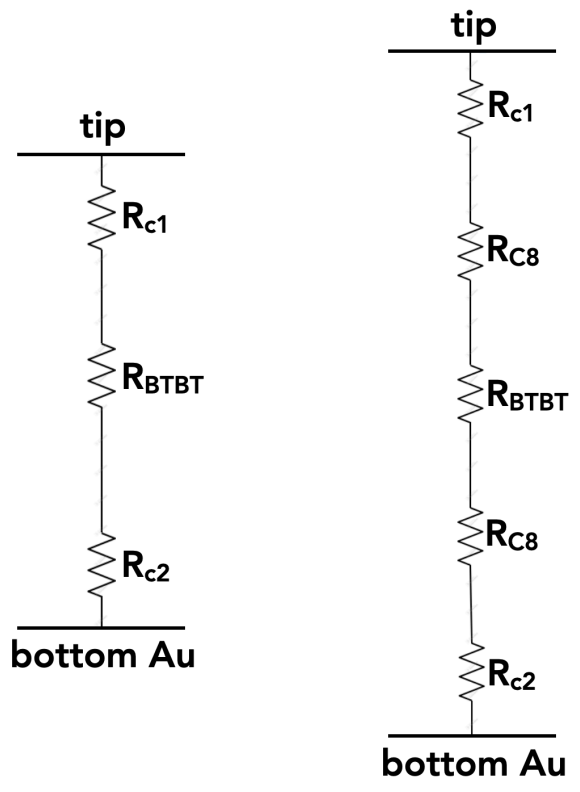

***Figure S6***. *Thermal equivalent circuit of $^{TS}$Au-S-C-BTBT MJ (left) the $^{TS}$Au-S-$C_8$-BTBT-$C_8$ MJ (right).*

## VIII.DFT calculations.

The geometric structure of the molecules was first optimized in the gas phase at the DFT level, with the B3LYP functional[38] and a 6-31 G (d,p) basis set[39] with the Gaussian09 software.[40] The molecules are tilted to fit the measured SAM thickness and then anchored on the gold (111) surface through a sulfur atom. The unit cell of the gold surface is modeled by a slab of three layers with 2 × 3 Au atoms in each layer and lattice parameters: a = 5.76 Å, b = 8.65 Å and α = 120° (see Fig. S7). With one molecule per unit cell, this corresponds to a theoretical density of 0.43 $nm^2$/molecule, which fits the experimental density deduced from the ellipsometry measurements (≈ 0.45 $nm^2$/molecule, see Tables S1). We consider here two tilted conformations for Au-S-C-BTBT-C interface referred to as "tilted1" and "tilted2"(see Fig. S7).

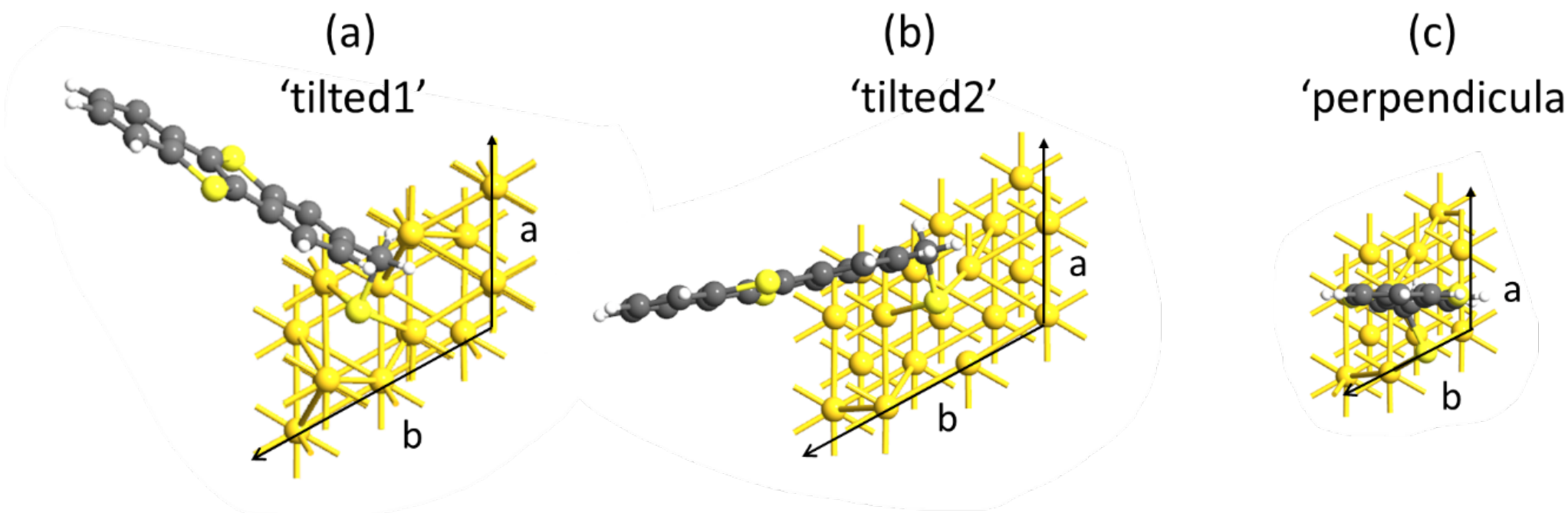

***Figure S7. T****op view of the optimized Au-S-C-BTBT interfaces in (a) tilted1, (b) tilted2 and (c) perpendicular conformations.*

The geometry of the interfaces was then optimized by relaxing the molecule and the top two gold layers until forces are below 0.01 eV/Å. The exchange and correlation effects are described by using the GGA.PBE exchange-correlation functional,[41] incorporating dispersion forces by Grimme correction (PBE+D2).[42] We expand the valence electrons in single zeta plus polarization (SZP) for gold atoms and double zeta polarization basis set (DZP) for the other atoms. The core electrons are frozen and described by the norm-conserving Troullier−Martins pseudopotentials.[43] A density mesh cutoff of 100 Ha and a (6×4×1) Monkhorst Pack k-sampling were used for the optimization. Once the geometry of the interface is optimized, a layer of gold ghost atoms has been added on the top layer of the gold electrodes at a distance of 1.7 Å away so that the work function of the clean Au (111) surface of 5.25 eV matches the experimental value[44] and previous theoretical studies.[45, 46] To build a single molecular junction, we add a second gold electrode on the top side of the molecular layer by assuming a van der Waals contact, with an interatomic distance determined as the sum of van der Waals radii of the hydrogen and gold atoms (2.86 Å) (see Fig. S8).

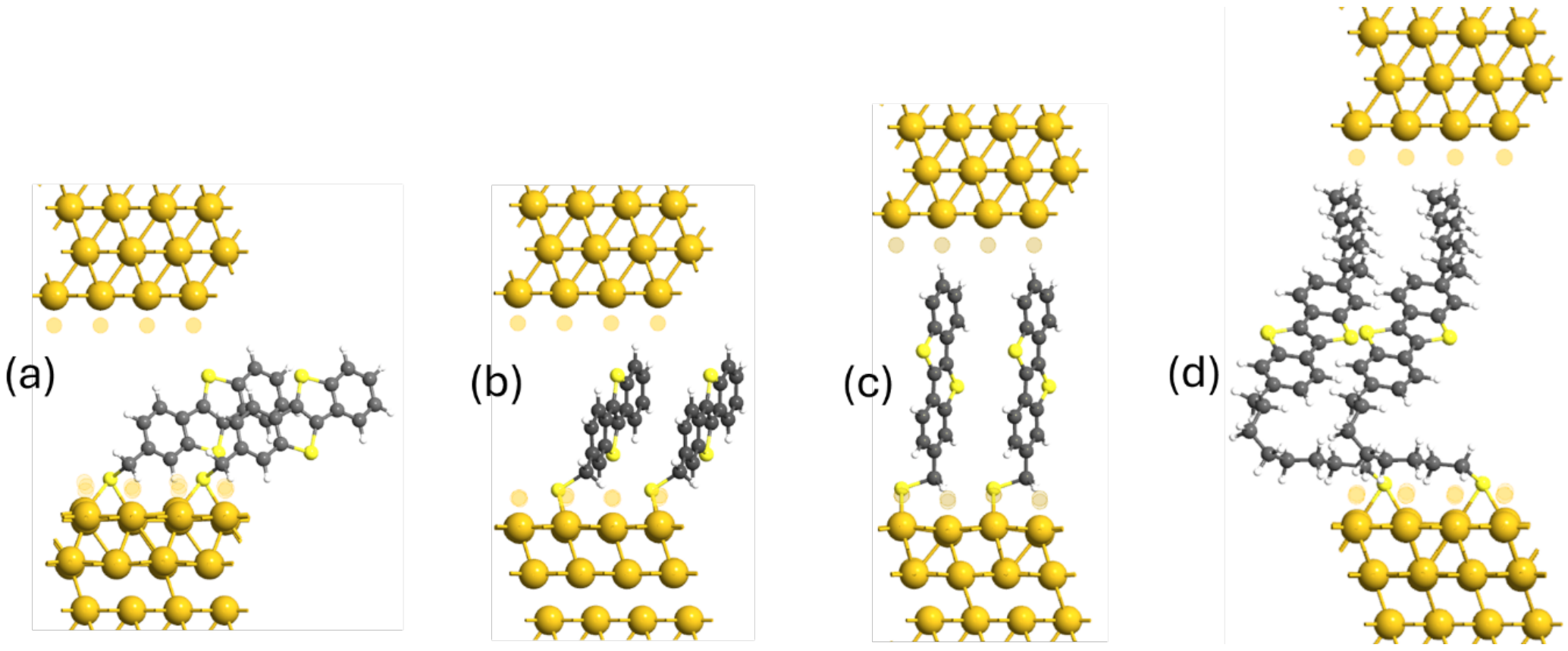

***Figure S8.*** *The optimized structure for both Au-S-C-BTBT/Au junction in (a) tilted1, (b) tilted2 and (c) perpendicular conformations and Au-S-$C_8$-BTBT-$C_8$/Au junction.*

The electronic transport calculations of the Au-S-C-BTBT/Au and Au-S-$C_8$-BTBT-$C_8$ junctions were performed by the combination of DFT to the Non-Equilibrium Green's Function (NEGF) formalism,[47] as implemented in QuantumATK Q-2023.12 package.[48-50] The exchange-correlation potential is described with the GGA.PBE functional[41] whereas the Brillouin zone was sampled with a (9×6×100) k-sampling. The mesh cutoff was set to 100 Ha with a temperature of 300 K. The transmission spectra at zero bias are then calculated with a 27×18×100 k points mesh. These parameters have been carefully tested to ensure the convergence of the transmission spectrum. Moreover, we consider a smaller gold surface with a 2×2 gold atom by a layer, with a=b=5.76 Å which corresponds to more densely packed molecules with a packing density of 0.26 $nm^2$/molecule (see Fig. S7-c). The same approach for optimization and transport calculations is used as described above except for k-point sampling for relaxation (6×6×1), transport calculations (9×9×100) and transmission (27×27×100).

# References.